\documentclass[a4paper,11pt]{article}
\usepackage{jheppub} % for details on the use of the package, please see the JINST-author-manual
\usepackage{amsmath,amssymb,amsfonts,stmaryrd}
\usepackage[dvipsnames]{xcolor}
\usepackage[normalem]{ulem}

\def\a{{\alpha}}
\def\b{{\beta}}

\def\m{{\mu}}
\def\n{{\nu}}
\def\r{{\rho}}
\def\s{{\sigma}}

\def\cc{\mbox{(c.c.)}}

\preprint{\texttt{KIAS-P26017}}

\title{On the integrability of root-Kerr probe dynamics}

\author[a]{Sungsoo Kim,}
\author[b]{Sangmin Lee}
\affiliation[a]{Department of Physics and Astronomy, Seoul National University, \\
1 Gwanak-ro, Gwanak-gu, Seoul 08826, Korea}
\affiliation[b]{School of Physics, Korea Institute for Advanced Study, \\
85 Hoegi-ro, Dongdaemun-gu, Seoul 02455, Korea}

\emailAdd{sooo4017@snu.ac.kr, 
sangminlee@kias.re.kr}

\abstract{
In the background of a Kerr-Newman black hole, the motion of a scalar particle is integrable by virtue of an extra conserved charge known as Carter charge. When the particle is endowed with spin, it is known that another conserved charge, the R\"udiger charge, maintains the integrability at least at low orders in the spin magnitude. We explore the extent of this integrability in a simpler model where both the source and the probe are root-Kerr particles, the non-gravitating limit of the Kerr-Newman black hole. At the leading order in the probe charge, the integrability holds to all orders in the spin magnitude if the interaction vertices of the probe are dictated by the Newman-Janis shift. At the second order in the probe charge, the integrability can be extended to the spin-squared order but begins to fail at the spin-cubic order. An argument based on asymptotic conservation suggests that it is impossible to restore the conservation at the spin-cubic order by a further deformation of the probe action. We compare our results with related observations for Kerr black holes with gravitational interactions.}

\begin{document}
\maketitle
\flushbottom

\newpage 
\section{Introduction}
\label{sec:intro}

The motion of a probe particle in the Schwarzschild metric is integrable much like the central-force problem in Newtonian mechanics. When the background geometry is replaced with a Kerr black hole solution, the symmetry is reduced and 
the integrability may seem lost. 
However, Carter \cite{Carter:1968rr} showed that the general Kerr-Newman solution admits an extra conserved charge 
which restores integrability. 

When the probe particle carries spin (angular momentum due to its own rotation), 
the phase space of the probe has at least one additional degree of freedom accounting for the spin orientation. 
To maintain integrability, an additional conserved charge is required. 
R\"udiger \cite{rudiger:1981rsa,rudiger:1983rsa} found such a charge.
Unlike the point particle probe, 
the integrability of the spinning probe is only approximate, 
and should be tested order by order in powers of spin. 
Initially, the integrability was verified at the first order in spin. 

Recent works \cite{Compere:2021kjz,Compere:2023alp,Ramond:2024ozy} extended the integrability to the quadratic order in spin, 
where the Mathisson-Papapetrou-Dixon (MPD) equation \cite{Mathisson:1937zz,Papapetrou:1951pa,Dixon:1970zza}, 
generalized to the spin-quadrupole order, 
was used to verify the conservation of the Carter and R\"udiger charges. 
The R\"udiger charge remains unmodified to this order, whereas the Carter charge receives corrections involving the spin tensor of the probe and the curvature tensor of the background spacetime. 

In a scattering setting, a weaker notion of conservation can be tested extensively. 
The classical scattering is governed by the radial action, or more broadly classical eikonal \cite{Gonzo:2024zxo,Kim:2024svw,Alessio:2025flu,Kim:2025olv,Kim:2025gis,Kim:2025sey}. The asymptotic conservation means that the asymptotic forms of the conserved charges, where the curvature terms are dropped, commute with the radial action under the Poisson-Dirac algebra. 
Ref.~\cite{Akpinar:2025tct} showed that the asymptotic conservation of the Carter and R\"udiger charges at the 1st post-Minkowskian order (1PM) holds for all orders in spin, even away from the probe limit. 
At 2PM, the conservation holds up to the quartic order in probe spin but breaks down at the quintic order. 

In general, it would be useful to know the maximal extent to which the conservation is valid. 
As a related question, once the \emph{asymptotic} conservation is verified to a certain approximation, 
it would be desirable to determine whether the strict \emph{local} conservation holds throughout the probe trajectory.

The purpose of this paper is to answer these questions for a simpler cousin of the Kerr-Newman black hole, 
known as the root-Kerr particle \cite{Arkani-Hamed:2019ymq}. The root-Kerr particle is not gravitating, 
but carries an infinite series of predetermined electric and magnetic multipole moments. 
Both the source and the probe are taken to be root-Kerr particles. 
To the leading order in the probe charge, which we call the 1st post-Lorentzian order (1PL) \cite{Bern:2023ccb}, 
we show that the conservation of the two charges holds to all orders in spin, 
as anticipated in \cite{Akpinar:2025tct}. 
At 2PL, however, the conservation can only be extended to the quadratic order in spin and not further. 
We consider the most general deformation of the world-line interaction vertices, as well as  
the most general modification of the charges, 
but it seems impossible to revive the charge conservation to the spin-cubic order. 

Our analysis of the root-Kerr probe dynamics is based on the recently formulated universal world-line model of massive spinning particles \cite{Kim:2026bqi}. The equation of motion (EoM) is derived in a Hamiltonian formulation. 
The covariant spin condition $S_{\m\n} p^\n = 0$ is imposed throughout. 
The resulting EoM generalizes the Bargmann-Michel-Telegdi (BMT) equation \cite{Bargmann:1959gz} with an infinite series of interaction vertices. 
At 1PL, the interaction vertices are completely fixed by the Newman-Janis (NJ) shift \cite{Newman:1965tw}. 
Following Guevara et al. \cite{Guevara:2020xjx} and subsequent work \cite{Kim:2023aff,Kim:2024grz}, 
we are elevating the NJ shift from a solution generating technique for the source to a dynamical principle 
to determine the world-line theory of the probe. 
At 2PL, no simple principle determines the interaction vertices a priori; 
we seek interactions that enhance the conservation of the charges as much as possible. 

In section~\ref{sec:root-kerr}, 
after a brief review of the root-Kerr particle as a static source, 
we present the dynamical equations for the root-Kerr probe following the general framework of \cite{Kim:2026bqi}. 
We show how the Newman-Janis shift completely determines the 1PL interaction vertices \cite{Kim:2024grz}, 
and discuss some 2PL vertices relevant for later sections.  
In section~\ref{sec:conservation}, we explain how to test the charge conservation in our world-line model. 
Geometric symmetries associated with Killing vectors lead to exact charge conservation; interaction vertices compatible with the symmetries can be added indefinitely without violating conservation. 
For the Carter and R\"udiger charges, 
the conservation depends on the peculiarity of the interaction vertices and the modification of the charges 
due to coupling to the background field. 

Section~\ref{sec:1PL} and \ref{sec:2PL} examine the conservation of the Carter and R\"udiger charges at 1PL and 2PL, respectively. The exact (all order in spin) conservation at 1PL is derived without much difficulty. 
The R\"udiger charge is not modified at this order, while the Carter charge receives a non-trivial correction. 
At 2PL, the conservation is more delicate. 
For the field-strength-squared interactions without more derivatives, the conservation at the spin-squared order 
uniquely selects a particular coupling. As we try to extend the conservation to the spin-cubic order, 
however, the conservation breaks down regardless of the couplings. 

In section~\ref{sec:asymptotic}, we turn to the asymptotic conservation. 
At 1PL, we confirm the observation of \cite{Akpinar:2025tct} that the asymptotic version of the exact conservation continues to hold even away from the probe limit. 
At 2PL, we show that even under the weaker notion of asymptotic conservation, 
it is impossible to restore the conservation at the spin-cubic order. 
We conclude the paper with some discussions in section~\ref{sec:discussion}. 

\paragraph{Conventions} 

Our conventions for the flat Minkowski metric and Levi-Civita tensor are 
\begin{align}
    \eta_{\mu\nu} = \operatorname{diag}(-,+,+,+)\,, 
    \quad 
    \epsilon_{0123} = +1 \,.
\end{align}
The Hodge dual is defined by
\begin{align}
   \tilde{F}_{\mu\nu} =  (* F)_{\mu \nu}=\frac{1}{2} \epsilon_{\mu \nu \rho \sigma} F^{\rho \sigma} \,.
\end{align}
The self-dual (SD,$+$) and the anti-self-dual (ASD,$-$) parts are defined by
\begin{align}
F^{ \pm}=\frac{1}{2}(F \mp i *F) \quad \Longrightarrow \quad * (F^{ \pm}) = \pm i(F^{ \pm}) \,.
\end{align}

\noindent
\textbf{Note added} 
While this work was being completed, we received ref.~\cite{deFirmian:2026mln} which also reported the charge conservation of the root-Kerr source-probe system up to the spin-squared and field-strength-squared order. 
Our results agree with theirs when the two overlap;  
see appendix~\ref{sec:dFV} for explicit comparison.
A notable difference is that we realized the all-order spin exponentiation at 1PL via the NJ shift.

\newpage 
\section{Root-Kerr: source and probe} 
\label{sec:root-kerr}

 The root-Kerr particle was introduced in \cite{Arkani-Hamed:2019ymq} as a ``single-copy" cousin of the Kerr black hole solution. Equivalently, it can be regarded as the $G\rightarrow 0$ limit of the Kerr-Newman black hole solution 
\cite{Lynden-Bell:2002dvr,Lynden-Bell:2004coe}. 
A root-Kerr particle is characterized by its electric charge $q$ and spin-length vector $\vec{a}$. It carries an infinite series of electric/magnetic multipole moments. The magnetic dipole moment exhibits $g=2$. 

\subsection{Source} 
\label{sec:source}

We review the field profile of a root-Kerr source following  \cite{Lynden-Bell:2002dvr,Lynden-Bell:2004coe,Vines:2017hyw,Chung:2019yfs}. 
When we place a static root-Kerr particle at the origin of $\mathbb{R}^3$ 
and choose its orientation such that $\vec{a} = (0,0,a)$, the field configuration is most succinctly described by 
a complex shift (Appell 1887): 
\begin{align}
    \vec{E} + i \vec{B} = - \nabla \Phi \,,
    \quad 
    \Phi = \frac{q}{4\pi (x^2+y^2+(z-ia)^2)^{1/2}} \,. 
    \label{Phi-complex}
\end{align}
Clearly, the fields far away from $\vec{R}=(x,y,z)=0$ satisfy Maxwell's equations in vacuum. 
By expanding $\Phi$ in $a/|\vec{R}| \ll 1$, we recover the electric/magnetic multipole moments. 

The complex-valued (3d) scalar potential $\Phi$ is often convenient, 
but cannot be identified with the true scalar potential $\phi = A^0 = -A_0$. The real part of the scalar potential is fine, 
\begin{align}
    \phi = \frac{1}{2}(\Phi + \Phi^*) = \cos(\vec{a}\cdot\nabla)\frac{q}{4\pi R} \,, 
    \quad R \equiv (\vec{R}\cdot \vec{R})^{1/2} = (x^2+y^2+z^2)^{1/2} \,.
    \label{phi-real}
\end{align}
It gives the static electric field as usual via $\vec{E} = - \nabla \phi$. 
For the magnetic field, we apply the Gilbert-Amp\`ere duality 
to switch from a scalar potential to a vector potential, 
\footnote{
We thank Joon-Hwi Kim for pointing out the crucial role of Gilbert-Amp\`ere
duality in the Newman-Janis algorithm. See \cite{Kim:2024mpy,Kim:2026opo} for related discussions and visual illustrations.
}
\begin{align}
    \vec{B} =-\nabla \phi_\mathrm{M}= -\nabla \frac{1}{2i}(\Phi - \Phi^*) \approx \nabla  \times \vec{A} \,.
    \label{phi-magnetic}
\end{align}
The weak equality sign $(\approx)$ indicates that a delta-function term irrelevant for long-distance physics has been discarded. The explicit form of the vector potential is 
\begin{align}
    \vec{A} = -(\vec{a}\times \nabla) \operatorname{sinc}(\vec{a}\cdot\nabla)\frac{q}{4\pi R} \,, 
    \quad 
    \operatorname{sinc}(w) = \frac{\sin{w}}{w} \,.
\end{align}

It is useful to introduce the Newman-Janis (NJ) shifted \cite{Newman:1965tw} position vectors, 
\begin{align}
   \vec{R} = (x,y,z) 
   \quad 
   \rightarrow 
   \quad 
   \vec{R}_\pm \equiv  (x,y, z \pm ia) \,,
   \quad 
   R_\pm = (\vec{R}_\pm \cdot \vec{R}_\pm)^{1/2} \,. 
\end{align}
The SD/ASD components of the field strength can be expressed as  
\begin{align}
\label{F-SD-ASD}
    F^\pm = -\frac{q}{8\pi} \frac{1}{(R_\mp^2)^{3/2}} \vec{R}_\mp \cdot  \left[ dt \wedge d\vec{R} \pm i *_3 d\vec{R}\right] \,.
\end{align}
where 
\begin{align}
    d\vec{R} = (dx^1, dx^3, dx^3) \,,
    \quad 
    *_3 d\vec{R} = (dx^2 \wedge dx^3, dx^3 \wedge dx^1, dx^1 \wedge dx^2) \,.
\end{align}
The Killing-Yano (KY) tensor is defined as 
\begin{align}
\begin{split}
     Y 
     &= \vec{a} \cdot (dt \wedge d\vec{R}) - \vec{R} \cdot (*_3 d\vec{R})\,, 
     \\
 \Longrightarrow\quad    \widetilde{Y} &= - \vec{R}\cdot (dt \wedge d\vec{R}) - \vec{a}\cdot (*_3 d\vec{R})\,, 
\end{split}
    \label{Y-KS}
\end{align}
Its SD and ASD components are \cite{Adamo:2023fbj,Kim:2024dxo,Kim:2024mpy}
\begin{align}
\label{Y-SD-ASD}
    Y^\pm = \pm \frac{i}{2} \vec{R}_\mp \cdot \left( dt \wedge d\vec{R} \pm i *_3 d\vec{R}\right) \,.
\end{align}

The tensors $F^\pm$ and $Y^\pm$ satisfy a simple proportionality, 
\begin{align}
\label{F-propto-Y}
    F^\pm = \frac{q}{4\pi} \frac{(\pm i)Y^\pm}{R_\mp^3} \,.
\end{align}
One of the many implications is that $Y$ and $F$ as matrices commute with each other, 
\begin{align}
    F^\m{}_\n Y^\n{}_\r = Y^\m{}_\n F^\n{}_\r \,. 
    \label{YF-commute}
\end{align}
The tensors also satisfy the relation, 
\begin{align}
        (F^\pm Y^\pm)^\mu{}_\nu = (Y^\pm F^\pm )^\mu{}_\nu = \frac{q}{16\pi} \frac{ (\mp i)}{R_\mp} \delta^\mu{}_\nu = \pm \frac{i}{2}  (\xi\cdot A^\pm) \delta^\mu{}_\nu\,. 
    \label{FY-pm-KS}
\end{align}
Here, $\xi^\mu\partial_\m = \partial_t$ is the time-translation Killing vector. 
The components of $Y$ are linear in Minkowski coordinates, so their first derivatives are constants, 
\begin{align}
    \partial_\mu Y_{\nu\rho} = \epsilon_{\mu\nu\rho\, 0} =   \epsilon_{\mu\nu\rho\sigma} \xi^\sigma\,.
    \label{dY-epsilon K}
\end{align}
All higher derivatives vanish trivially. 
It has been shown in \cite{Kim:2024mpy} that the
physical picture behind the facts derived above, such as Eqs. \eqref{F-SD-ASD}, \eqref{Y-SD-ASD}, \eqref{F-propto-Y},
is that the root-Kerr solution is a system of SD and ASD dyons at $\vec{x} = \pm
i\vec{a}$.

%%%%%%%%%%%%%%%%%%%%%%%%%%%%%%%%%%%%%%%%%%%%%%%%%%%%%
\subsection{Probe at 1PL: Newman-Janis shift}
\label{sec:probe-1PL}

The world-line model of a relativistic spinning particle coupled to an arbitrary electromagnetic or gravitational field 
had been a long-standing problem. A recent work \cite{Kim:2026bqi} offered a comprehensive and unifying solution to the problem. 
Here we review parts of \cite{Kim:2026bqi} that are directly relevant to this paper. 

There are many spinning particle models which differ in the variables to parametrize the spin degrees of freedom. A key observation of \cite{Kim:2026bqi} is that 
there exists a model-independent universal subspace of the phase space, parametrized by position $x^\mu$, momentum $p^\mu$ and (sign-flipped) spin-length vector $y^\mu= - a^\mu$. 

The Poisson bracket on the $(x,y,p)$ space is given by 
\begin{align}
   \begin{aligned}
\{\;\;,\;\; \}^\circ &= \eta^{\mu \nu} \frac{\partial}{\partial x^\mu} \wedge \frac{\partial}{\partial p^\nu}-\frac{y^\mu p^\nu+p^\mu y^\nu
}{p^2} \frac{\partial}{\partial x^\mu} \wedge \frac{\partial}{\partial y^\nu} \\
&\qquad -\frac{\epsilon^{\mu \nu}[y, p]}{2 p^2}\left(\frac{\partial}{\partial y^\mu} \wedge \frac{\partial}{\partial y^\nu}+\frac{\partial}{\partial x^\mu} \wedge \frac{\partial}{\partial x^\nu}\right) .
\end{aligned}
\label{PB-free-spin}
\end{align}
Equivalently, in terms of the complex-valued position coordinate, $z^\mu = x^\mu + iy^\mu$, 
\begin{align}
   \begin{aligned}
\{\;\;,\;\; \}^\circ &= \eta^{\mu \nu} \left( \frac{\partial}{\partial z^\mu} \wedge \frac{\partial}{\partial p^\nu}
+  \frac{\partial}{\partial \bar{z}^\mu} \wedge \frac{\partial}{\partial p^\nu} \right)\\
&\qquad -2\frac{\epsilon^{\mu \nu}[y, p]-i(y^\mu p^\nu +p^\mu y^\nu)
}{ p^2} \frac{\partial}{\partial z^\mu} \wedge \frac{\partial}{\partial \bar{z}^\nu} \, .
\end{aligned}
\label{PB-free-spin-zzbar}
\end{align}
Note that $\{z^\mu , z^\nu\}^\circ = 0 = \{\bar{z}^\mu , \bar{z}^\nu\}^\circ$. Following \cite{Kim:2023aff}, we will call it a``zigzag" property. 
The superscript circle in $\{ f, g\}^\circ$ indicates that the Poisson bracket describes a free theory.

The $(x,y,p)$ phase space is 12-dimensional, but the Poisson bracket has rank 10. The two ``null" coordinates are 
\begin{align}
    \{ f , y^2p^2\}^\circ = 0 \,,
    \quad 
    \{ f , y\cdot p\}^\circ = 0 
    \quad 
    \mbox{for arbitrary } f \,. 
\end{align}
The non-invertibility of the Poisson bracket will cause some minor technical difficulties. 

The (world-line) time evolution is generated by the usual Hamiltonian, 
\begin{align}
    H = \frac{\kappa^0}{2} (p^2 +m^2) \,.
\end{align}
Here, $\kappa^0$ is the Lagrange multiplier associated with the world-line reparametrization invariance. 
The mass-squared $m^2$ is in general a function of the spin-magnitude, $y^2p^2$, 
but for spin-magnitude-preserving interactions, including all interactions considered in this paper, 
treating $m^2$ as constant does not affect the EoM \cite{Kim:2026bqi}. 
The free EoM is simply 
\begin{align}
\left. \frac{df}{d\tau}\right|_\text{free} = \{ f, H \}^\circ \,, 
\end{align}
which gives, upon choosing $\kappa^0 =1/m$,  
\begin{align}
    \dot{x}^\m = \frac{p^\m}{m} \,,
    \quad 
    \dot{y}^\m = 0 \,,
    \quad
    \dot{p}^\m = 0 \,.
\end{align}

The interaction is incorporated via a symplectic deformation \cite{Kim:2026bqi}, 
\begin{align}
    \omega = \omega^\circ + \omega'  \,.
    \label{sympl-deform}
\end{align}
For a system in which $\omega^\circ$ is invertible, the Poisson bracket after the deformation would be 
\begin{align}
\label{poisson-deform}
\begin{split}
        \{ f , g \}   &= \{ f, g \}^\circ -  \{ f, \zeta^m\}^\circ \omega'_{mn} \{\zeta^n , g\} 
        \\
       &= \{ f, g \}^\circ - \{ f, \zeta^m\}^\circ \omega'_{mn} \{\zeta^n , g\}^\circ 
       \\
       &\qquad \qquad \qquad+  \{ f, \zeta^m\}^\circ \omega'_{mn} \{\zeta^n , \zeta^p \}^\circ  \omega'_{pq} \{ \zeta^p, g\}^\circ      +\cdots  \,. 
\end{split}
\end{align}
Here, $\zeta^m$ represent all coordinates of the phase space. 
We are working with a non-invertible Poisson bracket, which may be regarded as a Dirac bracket in each of the models considered in \cite{Kim:2026bqi}. We cannot obtain the interacting Poisson bracket by directly inverting \eqref{sympl-deform}, but it can be shown that the formula \eqref{poisson-deform} still holds with a non-maximal-rank Poisson/Dirac bracket.  

The two simplest interaction terms are the electric monopole and the magnetic dipole couplings. 
In our notation, they contribute 
\begin{align}
    \omega' = q A_\mu(x) dx^\mu + q \frac{g}{2} \tilde{F}_{\m\n} y^\m dx^\n \,.
\end{align}
The resulting EoM truncated at the $\mathcal{O}(q)$ order is the celebrated Bargmann-Michel-Telegdi (BMT) equation \cite{Bargmann:1959gz}, 
\begin{align}
\begin{aligned}
\dot{x}^\mu & =\frac{p^\mu}{m}-\frac{q}{m}\left[\frac{g}{2} \tilde{F}^\mu{}_\nu +\left(1-\frac{g}{2}\right) \hat{\delta}^\mu{}_\kappa  \tilde{F}^\kappa{}_\nu \right] y^\nu + \cdots, 
\\
\dot{y}^\mu & =\frac{q}{m}\left[F^\mu{}_\nu -\left(1-\frac{g}{2}\right) \hat{\delta}^\mu{}_\kappa F^\kappa{}_\n \right] y^\nu+\cdots, 
\\
\dot{p}^\mu & =\frac{q}{m} F^\mu{}_\nu p^\nu + \cdots,
\end{aligned}
\end{align}
where $\hat{\delta}^\m{}_\n = \delta^\m{}_\n - p^\m p_\n/p^2$ 
and $\tilde{F}_{\\m\n}$ is the Hodge dual of $F_{\m\n}$. 

\paragraph{Newman-Janis shift} 

For the root-Kerr particle, the $\mathcal{O}(q)$ part of the interaction term \eqref{sympl-deform} is completely fixed by the requirement that 
it reproduces the static solution \eqref{Phi-complex}. 
The resulting coupling is best summarized by the dynamical NJ shift \cite{Guevara:2020xjx,Kim:2023aff,Kim:2024grz}, which links the holomorphy in the complex spacetime to the self-duality of the field-strength \cite{Newman:1974fr,Kim:2023aff}, 
\begin{align}
    \omega'_\text{NJ} = \frac{q}{2} F_{\mu\nu}^+(z) dz^\mu \wedge dz^\nu +  \frac{q}{2} F_{\mu\nu}^-(\bar{z}) d\bar{z}^\mu \wedge d\bar{z}^\nu \,. 
\end{align}
The resulting EoM reads 
\begin{align}
\begin{aligned}
    \dot{p}_\mu &= q F^+_{\mu\nu}(z) \dot{z}^\nu + q F^-_{\mu\nu}(\bar{z})\dot{\bar{z}}^\nu \,,
    \\
    \dot{z}^\mu &= \frac{ p^\mu}{m} + \frac{2iq}{m^2} \left[ y^\mu p^\nu + p^\mu y^\nu  
    + i \epsilon^{\mu\nu\rho\sigma}y_\rho p_\sigma \right] F^-_{\nu\lambda} (\bar{z}) \dot{\bar{z}}^\lambda  \,,
    \\ 
    \dot{\bar{z}}^\mu &= \frac{ p^\mu}{m} - \frac{2iq}{m^2} \left[ y^\mu p^\nu + p^\mu y^\nu  
    - i \epsilon^{\mu\nu\rho\sigma}y_\rho p_\sigma \right] F^+_{\nu\lambda} ({z}) \dot{{z}}^\lambda \,.
\end{aligned}
\label{EoM-zzbar}
\end{align}
As shown in \eqref{poisson-deform}, the appearance of time-derivatives on the RHS is a generic feature 
of a symplectic perturbation theory. Iterating the EoM leads to an infinite series of terms on the RHS. 

%\newpage 
\subsection{Probe at 2PL: contact term}
\label{sec:probe-2PL}

Beyond the 1PL order, the Newman-Janis shift does not fix all couplings. 
There can be an infinite series of $\partial^k F^l$ couplings with undetermined coefficients. 
See \cite{Kim:2025xka} for an explicit and systematic discussion on this
point. 
We will be pragmatic and look for a coupling at 2PL which %simplifies the EoM and 
enhances the charge conservation. 

The EoM from the NJ shift coupling, before addition of further couplings, reads 
\begin{align}
\begin{aligned}
    \dot{p}_\mu &= q F^+_{\mu\nu}(z) \dot{z}^\nu + q F^-_{\mu\nu}(\bar{z})\dot{\bar{z}}^\nu \,,
    \\
    \dot{z}^\mu &= \frac{ p^\mu}{m} - 2iq M^{\m\n} F^-_{\nu\lambda} (\bar{z}) \dot{\bar{z}}^\lambda  \,,
\end{aligned}
\label{EoM-1PL-copy}
\end{align}
where 
\begin{align}
    M^{\mu\nu} =  \frac{1}{p^2}\left(y^\mu p^\nu + p^\mu y^\nu  
    + i \epsilon^{\mu\nu\alpha\beta}y_\alpha p_\beta \right)\,.
\end{align}
Iterating the EoM and extracting the $\mathcal{O}(q^2)$ terms, we get 
\begin{align}
\begin{aligned}
      \dot{p}_\m|_{q^2} &= -\frac{2iq^2}{m} (F^+MF^- p)_\m + \frac{2iq^2}{m} (F^- \bar{M} F^+ p)_\m \,,  
      \\
      \dot{z}^\m|_{q^2} &= \frac{4q^2}{m} (MF^- \bar{M}F^+p)^\m \,.
\end{aligned}
\end{align}
For brevity, we are using shorthand notations such as 
\begin{align}
    (ABCv)^\m = A^{\m\n} B_{\b\r} C^{\r\s} v_\s \,. 
\end{align}
To simplify the EoM, we use an identity, 
\begin{align}
    \bar{M} F^+ p = -  F^+ y \,,
    \qquad
    M F^- p = - F^- y \,.
    \label{MFp-identity}
\end{align}
The derivation of the identity is as follows. 
\begin{align}
\begin{split}
      p^2  \bar{M}^{\mu\nu} F^+_{\nu\rho} p^\rho &=   (y^\mu p^\nu + p^\mu y^\nu  
    - i \epsilon^{\mu\nu\alpha\beta} y_\alpha p_\beta)F^+_{\nu\rho} p^\rho
    \\
    &= (-y^\mu p^\nu + p^\mu y^\nu  
    - i \epsilon^{\mu\nu\alpha\beta} y_\alpha p_\beta)F^+_{\nu\rho} p^\rho
    \\
    &= - 2[(y\wedge p)^-]^{\mu\nu} F^+_{\nu\rho} p^\rho
    \\
    &= - 2 (F^+)^{\mu\nu}[(y\wedge p)^-]_{\nu\rho} p^\rho
     \\
    &= (F^+)^{\mu\nu}(-y_\nu p_\rho + p_\mu y_\nu  
    - i \epsilon_{\nu\rho}[y,p]) p^\rho 
    = - p^2  (F^+)^{\mu\nu} y_\nu \,.
\end{split}
\end{align}
After the simplification, the EoM takes the form 
\begin{align}
\begin{aligned}
      \dot{p}_\m|_{q^2,\text{undeformed}} &= 0  \,,  
      \\
      \dot{z}^\m|_{q^2,\text{undeformed}} &= - \frac{4q^2}{m} (MF^- F^+ y )^\m
      = \frac{4q^2}{m} \{ z^\m, y^\n \}^\circ (F^- F^+ y)_\n\,. 
\end{aligned}
\end{align}
The last expression in the second line is suggestive. 
By adding a suitable 2PL coupling, we can cancel the $q^2(F^+ F^-)$ terms completely, at the expense of introducing some field-derivative terms. 
We implement the cancellation using a Hamiltonian deformation, 
\begin{align}
    H_0 = \frac{\kappa}{2} (p^2+ m^2)
    \quad \rightarrow \quad 
    H_0 + q^2 H_2= \frac{\kappa}{2} \left[ p^2+ m^2 -4 q^2 yF^+(z)F^-(\bar{z}) y \right] \,. 
    \label{H-deformation}
\end{align}
The possibility of such a deformation was noted around eq.(5.31) of \cite{Kim:2024grz}. 
After adding the deformation, 
the final form of the $\mathcal{O}(q^2)$ EoM becomes 
\begin{align}
\label{EoM-full-deformed}
\begin{split}
\dot{p}^\mu|_{q^2,\text{deformed}} &=  \frac{q}{m} \left( F^+ + F^-\right)^\mu{}_\nu p^\nu   + \frac{2q^2}{m} (h^\m+\bar{h}^\m) + \mathcal{O}(q^3) \,, 
\\
    \dot{z}^\mu|_{q^2,\text{deformed}} &= \frac{p^\mu}{m}  + \frac{2iq}{m} (F^-)^\mu{}_\nu y^\nu -\frac{4iq^2}{m} M^{\mu\nu}  \bar{h}_\nu   + \mathcal{O}(q^3)  \,, 
\end{split}
\end{align}
where the vectors $(h_\m, \bar{h}_\m)$ carry the field-derivative effects, 
\begin{align}
    h_\mu = y (\partial_\mu F^+)F^-y \,,
    \quad 
    \bar{h}_\mu = y F^+ (\partial_\mu F^-)y \,.
    \label{h-bar-h}
\end{align}
We will work this version of the EoM in the rest of this paper.

We could implement the same cancellation using a symplectic deformation, 
\begin{align}
\label{2PL-deform-theta}
    \theta'_{(2)} = - \frac{2q^2}{m^2} (yF^+F^-y) p_\n dx^\n \,.
\end{align}
The Hamiltonian deformation \eqref{H-deformation} and the symplectic deformation \eqref{2PL-deform-theta} can be mapped to each other by a non-canonical redefinition of the momentum variable, if higher order corrections can be neglected.

%%%%%%%%%%%%
%\newpage 
\section{Conserved charges}
\label{sec:conservation}

\subsection{Generality}

A Hamiltonian system is defined by $(\omega, H)$, a symplectic form and a Hamiltonian, on a phase space. A flow $V = V^m \partial_m$ on the phase space is called a symmetry if 
\begin{align}
    \mathcal{L}_V \omega = 0 \,,
    \quad 
    \mathcal{L}_V H = V[H] = 0 \,.
\end{align}
Using Cartan's formula, we note that 
\begin{align}
     \mathcal{L}_V \omega = (d \,\mathrm{i}_V + \mathrm{i}_V \,d)\omega = d (\mathrm{i}_V\omega) = 0 \,.
     \label{diV-omega}
\end{align}
If the phase space is topologically trivial, the closed form $\mathrm{i}_V\omega$ should be exact:
\begin{align}
    \mathrm{i}_V\omega = - dQ \,.
    \label{V-omega-Q}
\end{align}
It follows that  $V$ is generated by $Q$:
\begin{align}
\label{V-generated-by-Q}
\begin{split}
     & V^m \omega_{mn} = - \omega_{nm} V^m = - \partial_n Q 
     \\
    \Longrightarrow 
    \qquad 
    &\delta_V f = V[f] =  V^m \partial_m f =  \omega^{mn} \partial_m f \partial_n Q = \{ f, Q\} \,.
\end{split}
\end{align}
Given a symplectic potential $\theta = \theta_n d\zeta^n$ ($\omega = d\theta$), 
$ \mathcal{L}_V \omega =0$ implies that 
\begin{align}
    \mathcal{L}_V \theta = (d \,\mathrm{i}_V + \mathrm{i}_V \,d)\theta = d\lambda 
    \quad \Longrightarrow \quad 
    d(\mathrm{i}_V \theta-\lambda) = - \mathrm{i}_V\omega 
    \quad \Longrightarrow \quad  
    Q= \mathrm{i}_V \theta-\lambda \,.
\end{align}

\paragraph{Constraints} 

The discussion above assumes a non-degenerate symplectic form and no constraints, so the components of the Poisson bi-vector is the matrix inverse of those of the symplectic form. 
When there are constraints, we work with the Dirac bracket with a non-maximal rank. 
The discussion of charge conservation is modified accordingly, as detailed in \cite{Kim:2026bqi}. 
A key observation is that, for some symmetries including the Killing symmetry to be reviewed below, the charge $Q$ and the Poisson bracket are both deformed by the interaction, 
while the vector field $V$ in \eqref{V-generated-by-Q} remains the same:
\begin{align}
\label{V-vs-Q}
    V[f] = \{ f, Q_0 \}^\circ =  \{ f, Q \} \,. 
\end{align}

\paragraph{Symplectic perturbation}

Consider a symplectic deformation with a coupling constant $q$, 
and the corresponding deformation of $V$ and $Q$. 
\begin{align}
    \begin{split}
        \omega &= \omega^\circ + q\, \omega' \,,
        \\
        V & = V_0 + q V_1 + q^2 V_2 + \cdots \,,
        \\
        Q &= Q_0 + q Q_1 + q^2 Q_2 + \cdots \,,
    \end{split}
\end{align}
Assuming $i_{V_0} \omega^\circ = - dQ_0$ 
and given $\omega'$, how could we construct $V_n$ and $Q_n$ $(n\ge 1)$? 

Expanding \eqref{diV-omega} and \eqref{V-omega-Q} in powers of $q$, we get 
\begin{align}
 d(\mathrm{i}_{V_{n-1}} \omega' +  \mathrm{i}_{V_{n}} \omega^\circ ) = 0 
 \quad 
 \Longrightarrow 
 \quad 
    \mathrm{i}_{V_{n-1}} \omega' +  \mathrm{i}_{V_{n}} \omega^\circ = -dQ_n \,.
    \label{V-Q-series}
\end{align}
In general, we may have to solve an infinite series of equations 
where the existence or uniqueness of the solution is not obvious. 
But when we can find $d(\mathrm{i}_{V_{k-1}} \omega') = 0$ for some $k$, we can set $V_{n\ge k} =0$, terminating the series. 
The simplest case is when $d(\mathrm{i}_{V_0} \omega') = 0$. 
Then $Q_1 = \mathrm{i}_{V_0} \theta' - \lambda_1$ and $Q = Q_0 + q Q_1$ is exactly conserved, 
while $V=V_0$ remains unchanged.  
This is the case with the Killing symmetry, in accordance with \eqref{V-vs-Q}.

\paragraph{Conservation with 1PL interaction}

Given a symplectic deformation at 1PL, the general EoM can be expressed as \cite{Kim:2024grz,Kim:2026bqi}
\begin{align}
    \frac{df}{d\tau} = \{ f, H\}^\circ - q \{ f, \zeta^m \}^\circ \omega'_{mn} \{\zeta^n, H\} \,.
\end{align}
The first two brackets on the RHS are the unperturbed ones while the last one is the full bracket. 
Applying it to the root-Kerr model with the NJ shift at 1PL (and no 2PL deformation yet) and substituting $Q_0$ for $f$, we get 
\begin{align}
\begin{split}
     \frac{d Q_0}{d\tau} &= \{ Q_0 , H \}^\circ -   q \sum_{a = \pm} \{ Q_0, z_a^\mu \}^\circ F^a_{\mu\nu}(z_a) \{ z_a^\nu, H\}
     \\
     &= q \sum_{a=\pm} \delta z_a^\mu  F^a_{\mu\nu}(z_a) \dot{z}_a^\nu \,.
\end{split}
\label{conservation-EoM}
\end{align}
Here, $z_+ = z$, $z_- = \bar{z}$. We are assuming the 0th order conservation $\{ Q_0 , H \}^\circ = 0$.

\subsection{Killing charges}

Suppose the background field admits a Killing vector $K = K^\mu \partial_\mu$ in the sense that 
\begin{align} \label{Killing_RK}
    \partial_\mu K_\nu + \partial_\nu K_\mu = 0 \,,
    \quad 
    (\mathcal{L}_K A)_\m = K^\nu \partial_\nu A_\mu + A_\nu \partial_\mu K^\nu = 0 \,.
\end{align}
In general, the requirement for $A_\m$ is that $\mathcal{L}_K A = d \lambda$ for some scalar $\lambda$. In flat Minkowski space without magnetic charges, one can always choose a gauge where $\lambda = 0$. We choose to work in such a gauge. 
In the spinless case ($y=0$), the conserved charge associated with the Killing vector is 
\begin{align}
    Q = K^\mu (p_\mu + q A_\mu) \,.
\end{align}

For a free spinning particle, described by the $(x,y,p)$ variables, 
the Killing charge generalizes to \cite{Kim:2026bqi} 
\begin{align}
    Q_0 = p_\mu K^\mu + \frac{1}{2} \epsilon^{\mu\nu\rho\sigma} y_\mu p_\nu \partial_\rho K_\sigma \,.
    \label{Q-free-spin}
\end{align}
When $K$ generates rotation, the two terms represent the orbital and spin angular momenta, respectively. 
For example, with  $K = x_1 \partial_2 - x_2 \partial_1$, we get
\begin{align}
    p_\mu K^\mu = x_1 p_2 - x_2 p_1 = L_{12} \,,
    \quad 
    \frac{1}{2} \epsilon^{\mu\nu\rho\sigma} y_\mu p_\nu \partial_\rho K_\sigma = -(y_0 p_3 - y_3 p_0) =  S_{12} \,.
\end{align}
Using the charge \eqref{Q-free-spin} and the bracket \eqref{PB-free-spin}, we can take the symmetry transformation of the variables, 
\begin{align}
\label{killing-variation-free}
\begin{split}
     \delta x^\mu  &= \{ x^\mu , Q_0 \}^\circ = K^\m(x) \,,
    \\ 
    \delta y^\mu &=  \{ y^\mu , Q_0 \}^\circ = -(\partial^\m K_\n) y^\n  \,, 
    \\
     \delta p^\mu  &= \{ p^\mu , Q_0 \}^\circ = -(\partial^\m K_\n) p^\n  \,.
\end{split}
\end{align}

\paragraph{Killing charges with 1PL interaction}

In flat Minkowski space, 
since $K^\mu(x)$ is only a linear function of $x$, we have
\begin{align}
    \delta z^\mu = \delta x^\mu + i \delta y^\mu = K^\mu(x) + i y^\nu \partial_\nu K^\mu = K^\mu(x+iy) = K^\mu(z) \,.
\end{align}
The RHS of \eqref{conservation-EoM} becomes 
\begin{align}
    q \sum_{a = \pm} K^\mu(z_a) F^a_{\mu\nu}(z_a) \dot{z}_a^\nu = q \sum_{a = \pm} \left[ K^\mu (\partial_\mu A_\nu - \partial_\nu A_\mu)\dot{z}^\nu\right]_a \,.
\end{align}
The Killing condition for $A_\mu$ implies that 
\begin{align}
\begin{split}
      K^\mu (\partial_\mu A_\nu - \partial_\nu A_\mu)\dot{z}^\nu &= - (A_\mu \partial_\nu K^\mu + K^\mu \partial_\nu A_\mu)\dot{z}^\nu
      \\
      &= -\dot{z}^\nu \partial_\nu (K^\mu A_\mu) = - \frac{d}{d\tau}(K^\mu A_\mu) \,.  
\end{split}
\end{align}
Therefore we conclude that
\begin{align}
    \frac{d}{d\tau}(Q_0 + q Q_1) = 0 \,,
    \quad Q_1 = K^\mu(z)A_\mu^+(z) +  K^\mu(\bar{z})A_\mu^-(\bar{z}) \,.
    \label{Killing-root-Kerr}
\end{align}
In summary, the expression for a conserved momentum associated with a Killing vector is 
\begin{align}
    Q = K^\mu(x) p_\mu + \frac{1}{2} \epsilon^{\mu\nu\rho\sigma} y_\mu p_\nu \partial_\rho K_\sigma+ q \left[ K^\mu(z)A_\mu^+(z) +  K^\mu(\bar{z})A_\mu^-(\bar{z}) \right] \,.
    \label{Killing-momentum-final}
\end{align}
The modification of the charge due to the interaction is such that 
the flow-charge relation \eqref{V-vs-Q} holds. 

\paragraph{Killing charges with 2PL interaction} 

If we were to add the 2PL interaction through symplectic deformation, as illustrated in \eqref{2PL-deform-theta}, 
the Killing charge $Q$ will receive a further correction. But, if we add the 2PL interaction by the Hamiltonian deformation as in \eqref{H-deformation}, the form of the charge does not change. 
To preserve the symmetry, we only need to check 
\begin{align}
   \{ Q_0 + q Q_1, H_0 + q^2 H_2\} = 0 \,.  
\end{align} 
Since $H_2$ depends on the field strength produced by the source, it will preserve precisely those symmetries left unbroken by the source.

%%%
%\newpage 
\subsection{Carter and R\"udiger charges}

We introduce the Carter \cite{Carter:1968rr} and R\"udiger \cite{rudiger:1981rsa,rudiger:1983rsa} charges 
and discuss briefly how the probe spin modifies the definitions. 
See {\it e.g.} refs.~\cite {Gibbons:1993ap,Rosquist:2007uw,Will:2008ys} for further discussions on these charges. 

For a spinless probe $(y^\m=0)$, in terms of the Killing-Yano tensor \eqref{Y-KS}, the Carter charge is defined as
\begin{align}
    C_0|_\mathrm{spinless} = Y_\mu Y^\mu = K_{\rho \sigma} p^\rho p^\sigma \,,
    \quad 
    Y_\mu \equiv Y_{\mu\rho} p^\rho \,,
    \quad 
    K_{\rho\sigma} = Y_{\rho\mu} Y_\sigma{}^\mu \,. 
\end{align}
Applying the EoM for a spinless probe, we find 
\begin{align}
\begin{split}
     \left. \frac{d C_0}{d\tau}\right|_\mathrm{spinless} &= -2 p\cdot Y\cdot Y \cdot \frac{dp}{d\tau}  
      = -\frac{2q}{m} (p\cdot Y\cdot Y\cdot F\cdot p) \,.
\end{split}
\label{K-Y-spinless}
\end{align}
%If $Y$ and $F$ are arbitrary 2-forms, the RHS has no reason to vanish. But, 
As we noted in \eqref{YF-commute}, 
the two matrices commute $Y\cdot F = F \cdot Y$, 
so the RHS of does vanish thanks to the antisymmetry of $Y\cdot F\cdot Y$. 

The Carter charge was generalized to include spin by R\"udiger \cite{rudiger:1981rsa,rudiger:1983rsa} and further extended by Comp\`ere et al.~\cite{Compere:2021kjz,Compere:2023alp}. In the root-Kerr limit, where the curvature terms vanish, 
\begin{align}
\begin{split}
     C_0 &= (Y_{\nu\rho}  p^\rho) (Y^{\nu\sigma}p_\sigma) - 4 \xi^\lambda \epsilon_{\lambda \mu \sigma [\rho} Y_{\nu]}{}^\sigma S^{\mu\nu} p^\rho 
     + \eta_{\mu\nu} \left[ \xi_\rho \xi_\sigma -\frac{1}{2} \eta_{\rho\sigma}\xi^2\right] S^{\mu\rho} S^{\nu\sigma} \,.   
\end{split}
\end{align}
Our definition differs from that of ref.~\cite{Compere:2023alp} by an overall sign. 
In what follows, to avoid clutter, we use abbreviations such as $xy = x\cdot y$, $x A y = x\cdot A \cdot y$, 
and $w_\mu = \epsilon_\mu[y,p,\xi]$.  
After the substitution, $S^{\mu\nu} = \epsilon^{\mu\nu}[y,p]$, 
we can express $C_0$ as 
\begin{align}
\label{C0-flat}
\begin{split}
    C_0 &=  (Y_{\nu\rho}  p^\rho)^2
    +2 p^2 (yY \xi) + 2 (\xi p)(y Y p) +w^2  + \xi^2  y^2 p^2 
    \\
     &= - pYYp
    +2 p^2 (yY \xi) + 2 (\xi p)(y Y p) + p^2 (\xi y)^2  + y^2 (\xi p)^2   \,.  
\end{split}
\end{align}

The R\"udiger charge is linear in the KY tensor: 
\begin{align}
   R_{0} = Y_{\mu\nu} y^\mu p^\nu \,.
\end{align}
The conservation for a spinless probe follows easily, 
\begin{align}
    \frac{dR_0}{d\tau} = \frac{q}{m}\left( - y \cdot F\cdot Y\cdot p + y \cdot Y \cdot F\cdot p \right) = 0 \,, 
\end{align}
where we used the commutativity $Y\cdot F = F\cdot Y$ again. 

In the gravitational setup, where both the source and the probe are Kerr black holes, ref.~\cite{Compere:2023alp} showed that the R\"udiger charge continues to be conserved without any modification up to the spin-squared order. 
In the following two sections, we will examine the conservation of $C$ and $R$ to all orders in spin at 1PL and to the spin-squared order at 2PL, while adding correction terms to the charges if necessary.

%%%%%%%%%%%%%%%%%%%%%%%%%%%%%%%%%%%%
%\newpage 
\section{Exact conservation at 1PL} 
%%%%%%%%%%%%%%%%%%%%%%%%%%%%%%%%%%%%
\label{sec:1PL}

At the 1PL order, assuming that the EoM is completely dictated by the NJ shift, we will show that both the Carter charge and the R\"udiger charge are conserved to all orders in spin. 
The definition of the R\"udiger charge is not modified, whereas the Carter charge receives a non-trivial correction. 

\subsection{R\"udiger charge}

Following the general discussion in the previous section, we begin by computing the symmetry variation of the dynamical variables:
$\delta f = \{ f, R_0 \}^\circ$ before turning on the EM background. 
First, 
\begin{align}
      \delta p^\mu &= -\partial^\mu Y_{\rho\sigma} y^\rho p^\sigma = - \epsilon^\mu[y,p,\xi] \,.
      \label{delta-p-Rudiger}
\end{align}
To compute $\delta z^\mu$, we note that $Y$ is linear in $x$, which implies 
\begin{align}
    Y_{\nu \sigma}(z) - Y_{\nu\sigma}(x) = iy^\rho \partial_\rho Y_{\nu\sigma} = iy^\rho \epsilon_{\rho\nu \sigma \alpha} \xi^\alpha \,.
\end{align}
It follows that 
\begin{align}
    R_{0} = y^\nu Y_{\nu\sigma}(x) p^\sigma = y^\nu Y_{\nu\sigma}(z) p^\sigma  =y^\nu Y_{\nu\sigma}(\bar{z}) p^\sigma \,.
\end{align}
Using the zigzag property, %% introduce it in the root-Kerr review 
we find 
\begin{align}
\begin{split}
      \delta z^\mu &= \{ z^\mu, R_0\} 
      = \{ z^\mu, y^\nu Y_{\nu\sigma}(z) p^\sigma \} 
      \\
      &= y_\nu Y^{\nu\mu}(z) + \{z^\mu, y^\nu \} Y_{\nu\sigma}(z) p^\sigma
      \\
      &= y_\nu Y^{\nu\mu}(z) - \frac{1}{p^2} \left( y^\mu p^\nu +  p^\mu y^\nu  + i \epsilon^{\mu\nu}[y,p] \right) Y_{\nu\sigma}(z) p^\sigma
      \\
      &= -\frac{1}{p^2} \epsilon^{\mu\nu}[y,p] (\tilde{Y}_{\nu\sigma}(z) + i Y_{\nu\sigma}(z))p^\sigma 
      = -\frac{2i}{p^2} \epsilon^{\mu\nu}[y,p] Y^+_{\nu\sigma}(z) p^\sigma 
    \\
    &= - 2\left(\eta^{\mu\nu} - \frac{p^\mu p^\nu}{p^2} \right) Y^+_{\nu\rho}(z) y^\rho 
    = 2 y^\rho Y^+_{\rho\nu}(z) \left(\eta^{\nu\mu} - \frac{p^\nu p^\mu}{p^2} \right) \,.
\end{split}
\label{delta-z-rudiger}
\end{align}
A correlation between self-duality and holomorphy has emerged.

Now we can run the EoM test  \eqref{conservation-EoM} of the conservation of the R\"udiger charge. 
It follows from \eqref{delta-z-rudiger} that 
\begin{align}
    \left. \frac{dR_0}{d\tau}\right|_{q^1} = \frac{2q}{m} \sum_{a=\pm} y^\rho Y^a_{\rho\sigma}(z_a) \hat{\eta}^{\sigma \mu} F^a_{\mu\nu}(z_a) p^\nu \,, 
    \qquad 
    \hat{\eta}^{\sigma\mu} = \eta^{\sigma\mu} - \frac{p^\sigma p^\mu}{p^2} \,.
    \label{Q0-dot-rudiger}
\end{align} 
Using \eqref{FY-pm-KS}, we can show that the RHS of \eqref{Q0-dot-rudiger} vanishes:
\begin{align}
     \left.  \frac{dR_0}{d\tau} \right|_{q^1} = 0 \,.
\end{align}
In view of \eqref{V-Q-series}, we have shown that $V_1^m = -(\Pi^\circ)^{mn} \omega'_{nk}V_0^k$, $V_1[H]=0$ and $R_1=0$.

\subsection{Carter charge}

Our next goal is to check the 1PL conservation of the Carter charge, 
\begin{align}
\label{C0-copy}
\begin{split}
    C_0 &=  (Y_{\nu\rho}  p^\rho)^2
    +2 p^2 (yY \xi) + 2 (\xi p)(y Y p) +w^2  + \xi^2  y^2 p^2 
    \\
     &= - pYYp
    +2 p^2 (yY \xi) + 2 (\xi p)(y Y p) + p^2 (\xi y)^2  + y^2 (\xi p)^2   \,.  
\end{split}
\end{align}
The last term in the first line can be discarded, since it commutes with everything.
The $w^2$ term can be absorbed into the NJ shift of the first term, 
\begin{align}
    (Y_{\nu\rho}(x)  p^\rho) (Y^{\nu\sigma}(x)p_\sigma) + w^2 =  (Y_{\nu\rho}(z)  p^\rho) (Y^{\nu\sigma}(\bar{z})p_\sigma) \,. 
\end{align}
So we may work with a modified Carter charge, 
\begin{align}
    \widetilde{C}_0 &=  Y_{\nu\rho}(z)  p^\rho Y^{\nu\sigma}(\bar{z}) p_\sigma 
    +2 p^2 [yY(x)\xi] + 2 (\xi p)[y Y(x)p] \,. 
\end{align}
To compute $\delta z^\mu$, it is useful to rewrite $\widetilde{C}_0$ as 
\begin{align}
     \widetilde{C}_0 &=  Y_{\nu\rho}(z)  p^\rho Y^{\nu\sigma}(z) p_\sigma 
    + 4 p^2 [yY^+(z) \xi ] + 4 (\xi p)[y Y^-(z) p] \,. 
\end{align}
The zigzag property, $\{ z^\mu , z^\nu \} =0$, ensures that no derivative of $Y$, $\tilde{Y}$, $Y^\pm$ is ever produced during the computation of $\delta z^\mu = \{ z^\mu , \widetilde{C}_0 \}$. 
After some algebra, we can show that 
\begin{align}
\begin{split}
     \delta z^\mu &=  -2Y^{\mu\nu}(z) Y_{\nu\rho}(z) p^\rho  - 4 \xi^\mu \left( [y Y^+(z)p] - [y Y^-(z) p]\right)
     \\
     &\qquad+ 4p^\mu  [y Y^+(z) \xi] -4 y^\mu [p Y^+(z) \xi]
     +  4 Y^+(z)^{\mu\nu} \left[ (\xi y)  p_\nu - (\xi p) y_\nu \right]
     \\
     &= -2Y^{\mu\nu}(z) Y_{\nu\rho}(z) p^\rho + 4 \xi^\mu  [y Y^-(z) p] 
     - 4\left( y^\mu p^\nu - p^\mu y^\nu + i \epsilon^{\mu\nu}[y,p]\right) Y^+_{\nu\rho}(z) \xi^\rho 
     \\
     &=-2Y^{\mu\nu}(z) Y_{\nu\rho}(z) p^\rho + 4 \xi^\mu  [y Y^-(z) p] 
     +  4Y^+(z)^{\mu\nu} \left[ (\xi y)  p_\nu - (\xi p)  y_\nu -i  w_\nu \right] \,.
\end{split}
\label{dz-Cater-5}
\end{align}

%\newpage
\paragraph{EoM test - 1st order}

The EoM test \eqref{conservation-EoM} at 1PL gives
\begin{align}
\begin{split}
     \left. \frac{d C_0}{d\tau}\right|_{q^1} &= \frac{q}{m} \sum_{a=\pm} \delta z_a^\mu  F^a_{\mu\nu}(z_a) p^\nu \,.
\end{split}
\label{EoM-Carter-q1}
\end{align}
Many terms vanish or cancel out, but some survive. 
The problematic terms are 
\begin{align}
    (\delta z^\mu)_\mathrm{prob} = 4\xi^\mu[y Y^-(z)p] + 4(\xi y) Y^+(z)^{\mu\nu} p_\nu \,.
\end{align}
They lead to 
\begin{align}
    (\delta z^\mu)_\mathrm{prob} F_{\mu\nu}^+(z) p^\nu = 4 [\xi F^+(z)p][y Y^-(z) p] - 4 (\xi y)[pY^+(z)F^+(z)p] \,.
\end{align}
Interestingly, we can turn the RHS into a total derivative, 
using the following identities:
\begin{align}
\begin{split}
     \xi F^+(z)p &= - m \frac{d}{d\tau}[\xi A^+(z)] + \mathcal{O}(q) \,, 
     \\
   m  \frac{d}{d\tau} [y Y^-(z) p]&= m y^\nu \frac{d}{d\tau} \left[ Y^-_{\nu\sigma}(z)\right] p^\sigma + \mathcal{O}(q)
    \\
    &= - \frac{i}{2} y^\nu  (p_\nu \xi_\sigma - \xi_\nu p_\sigma) p^\sigma + \mathcal{O}(q)  
    %\\
    = \frac{i}{2} p^2(\xi y) + \mathcal{O}(q)  \,,
    \\
    p Y^+(z)F^+(z) p &= \frac{i}{2} p^2 [\xi A^+(z)] \,.
\end{split}
\end{align}
It follows that the Carter charge with a correction term is conserved at 1PL, 
\begin{align}
\begin{split}
        &\left. \frac{d}{d\tau}(C_0 + q C_1)\right|_{q^1} = 0 \,,
    \\
     &\qquad C_1 = 4 [\xi A^+(z)][ y Y^-(z) p] +  4 [\xi A^-(\bar{z})][ y Y^+(\bar{z}) p] \,.
\end{split}
\label{Carter-q1-final}
\end{align}
We can rewrite $C_1$ as 
\begin{align}
\begin{split}
    C_1 
    &= 2 \left[ \xi A^+(z) + \xi A^-(\bar{z})\right] \left( [y Y(x)p] + y^2 (\xi p) \right)
    \\
   &\qquad + 2i \left[ \xi A^+(z) - \xi A^-(\bar{z}) \right] [y \tilde{Y}(x)p]\,.    
\end{split}
\label{C1-alt}
\end{align}

\paragraph{Commutativity of the charges} 

In the absence of the EM interaction, 
one can show that the R\"udiger charge and the Carter charge Poisson-commute, 
\begin{align}
    \{ R_0 , C_0 \}^\circ = 0 \,.
    \label{R-C-commute-0}
\end{align}
We should check whether it continues to hold after the interaction is turned on. 
At 1PL, we need to verify that 
\begin{align}
    \{ R_0, C_1 \}^\circ -  \sum_{a=\pm} \{R_0, z^\mu_a \}^\circ F^a_{\mu\nu}(z_a) \{z_a^\nu , C_0\}^\circ = 0 \,.
        \label{R-C-commute-1a}
\end{align}
It is convenient to rewrite it as 
\begin{align}
    \sum_{a = \pm} \delta_R z_a^\mu\left( \frac{\partial C_1}{\partial z_a^\mu} - F^a_{\mu\nu}(z_a) \delta_C z_a^\nu \right) =  -\frac{\partial C_1}{\partial p^\mu} \delta_R p^\mu \,,
     \label{R-C-commute-1b}
\end{align}
where 
\begin{align}
    \delta_R f = \{f , R_0\}^\circ \,,\quad 
    \delta_C f = \{f , C_0\}^\circ \,. 
\end{align}
Given \eqref{delta-p-Rudiger}, the RHS of \eqref{R-C-commute-1b} is fairly simple, 
\begin{align}
   -\frac{\partial C_1}{\partial p^\mu} \delta_R p^\mu 
   = \frac{\partial C_1}{\partial p^\mu} w_\mu = 4[\xi A^+(z)] [y Y^-(z)w] +4[\xi A^-(\bar{z})][y Y^+(\bar{z})w] \,. 
\end{align}
On the LHS of \eqref{R-C-commute-1b}, a lot of cancellations occur before the contraction with $(\delta_R z^\mu)$:
\begin{align}
\begin{split}
     \frac{\partial C_1}{\partial z_\pm^\mu} - F^\pm_{\mu\nu}(z_\pm) \delta_C z_\pm^\nu &= 
     \pm i (\xi A^\pm)Y^\pm_{\mu\nu}(z^\pm) p^\nu
      +2 [F^\pm Y^\mp(z_\pm)Y^\mp(z_\pm)]_{\mu\nu} p^\nu 
     \\
     &\quad 
     \mp 2i (\xi A^\mp) Y^\pm_{\mu\nu}(z_\mp)p^\nu \,.  
\end{split}
\label{dC1-identity}
\end{align}
Upon the contraction with $(\delta_R z^\mu)$, only the last term gives a non-vanishing contribution. 
The remaining step to verify \eqref{R-C-commute-1b} is relatively simple.
The identity \eqref{dC1-identity} can be used for other purposes. For example, it can quickly double-check  \eqref{Carter-q1-final}.

%%%%%%
%\newpage 
\section{Approximate conservation at 2PL}
\label{sec:2PL}

\subsection{R\"udiger charge}

Suppose we implement the 1PL interaction governed by the NJ shift and (tentativley) do not include 
the 2PL deformation in \eqref{H-deformation} yet. 
The EoM test at 2PL would give  
\begin{align}
\begin{split}
     \left.  \frac{dR_0}{d\tau} \right|_{q^2,\text{undeformed}} &= \frac{4iq^2}{m} y^\rho\left[ Y^+_{\rho\sigma} \hat{\eta}^{\sigma\mu}(F^+ F^-)_{\mu\nu} - Y^-_{\rho\sigma} \hat{\eta}^{\sigma\mu}(F^- F^+)_{\mu\nu} \right] y^\nu   
     \\
     &= \frac{4iq^2}{m} y^\rho\left[ Y^+_{\rho\sigma} - Y^-_{\rho\sigma}  \right]\hat{\eta}^{\sigma\mu}(F^+ F^-)_{\mu\nu}   y^\nu \,.  
\end{split}
\end{align}
The relation \eqref{F-propto-Y} $F_\pm \propto Y_\pm$ implies that 
\begin{align}
    Y^+ F^+ F^- \propto Y^+ Y^- Y^+ 
    \quad \Longrightarrow \quad 
    y( Y^+ F^+ F^-)y \propto y(Y^+ Y^- Y^+)y = 0 \,.
\end{align}
So, we would be left with 
\begin{align}
\begin{split}
     \left.  \frac{dR_0}{d\tau} \right|_{q^2} 
     &= -\frac{4iq^2}{mp^2} y^\rho( Y^+_{\rho\sigma} - Y^-_{\rho\sigma} )p^\sigma p^\mu(F^+ F^-)_{\mu\nu}   y^\nu 
     \\
     &= -\frac{4q^2}{mp^2} [ y \tilde{Y}(x)  p ] \left[ p  F^+(z) F^-(\bar{z})    y \right] \,.  
\end{split}
\label{Rudiger-q2-1} 
\end{align}
It is not clear whether/how one can turn the RHS to a total derivative and absorb into the definition of the charge. 
As it stands, the RHS signals the violation of the charge conservation at the $\mathcal{O}(q^2y^2)$ order. 

The situation is improved slightly if we include the 2PL deformation in \eqref{H-deformation}. The EoM test now gives 
\begin{align}
    \left.  \frac{dR_0}{d\tau} \right|_{q^2,\mathrm{deformed}} = -\frac{2q^2 }{m}   y^\mu\{R_0 , [F^+(z)F^-(\bar{z})]_{\mu\nu} \} y^\nu \,.
    \label{Rudiger-q2-2} 
\end{align}
We can try to turn the RHS into total derivatives up to $\mathcal{O}(q^3)$ terms; 
\eqref{Rudiger-q2-2} implies that 
\begin{align}
\begin{split}
     \frac{d}{d\tau} \left( R_0 + q^2 R_2 \right) &= \frac{2iq^2}{p^2} [y \tilde{Y}(x) p]\left[ y \left(\frac{dF^+}{d\tau}\right) F^- y - y F^+ \left( \frac{dF^-}{d\tau}\right) y \right] + \mathcal{O}(q^3)
    \\ 
    &= \frac{2iq^2}{mp^2} [y \tilde{Y}(x) p] p^\mu (h_\mu - \bar{h}_\mu) + \mathcal{O}(q^3) \,,
\end{split}
    \label{Rudiger-q2-3} 
\end{align}
where $(h,\bar{h})$ are as defined in \eqref{h-bar-h}, and the correction to the charge is 
\begin{align}
   R_2 = 
 \frac{2}{p^2} \left[ yY(x)p - y^2(\xi p) \right]  [y F^+(z)F^-(\bar{z})y] \,. 
 \label{Rudiger-q2-correction}
\end{align}
In showing \eqref{Rudiger-q2-3}, we find some non-trivial vanishing of a group of terms: 
\begin{align}
\begin{split}
    & y^\alpha Y^+_{\alpha \beta}(z)   h^\b    +  y^\alpha Y^-_{\alpha \beta}(\bar{z})  \bar{h}^\b = 0 \,.
\end{split}
\end{align}

The conclusion drawn from \eqref{Rudiger-q2-3} is that the conservation of $R$ can be maintained up to the $q^2y^2$ order, 
but it breaks down at the $q^2y^3$ order. 
One may wonder whether it is possible to deform the theory further and absorb the RHS of \eqref{Rudiger-q2-3} into the definition of the charge. In section~\ref{sec:asymptotic}, we will argue that such a deformation is impossible.

%%%%%%%%%%%%%%%%%%%%%%%%%%%%%%%%%%%%%%%%%
%\newpage 
\subsection{Carter chage} 
%%%%%%%%%%%%%%%%%%%%%%%%%%%%%%%%%%%%%%%%%

Our findings regarding the R\"udiger charge suggests that 
we can aim for 
\begin{align}
\label{Carter-to-quadratic}
    \frac{d}{d\tau} (C_0 + q C_1 + q^2 C_2) = \mathcal{O}(q^2y^3) \,. 
\end{align}
The relevant EoM is \eqref{EoM-full-deformed}, which we reproduce here, 
\begin{align}
    \begin{aligned}
\dot{z}^\mu &= \frac{p^\mu}{m}  + \frac{2iq}{m} (F^-)^\mu{}_\nu y^\nu  + \mathcal{O}(y^3)  \,, 
\\
\dot{p}^\mu &=  \frac{q}{m} \left( F^+ + F^-\right)^\mu{}_\nu p^\nu  + \frac{2q^2}{m} (h^\mu +\bar{h}^\mu ) + \mathcal{O}(y^3)  \,. 
\end{aligned}
\label{EoM-full-deformed-y2}
\end{align}
It is convenient to separate the (energy)$\times$(R\"udiger) terms and define an effective charge. 
We should expand $C_1$ defined in \eqref{C1-alt} in powers of $y$ and collect all terms up to $\mathcal{O}(y^2)$. 
The effective charge thus constructed is 
\begin{align}
    C_\mathrm{eff} = -pYYp + 2 q \left[ (\xi \tilde{A}+\xi Fy) (y \tilde{Y}p) + y^2(\xi A) (\xi p)\right]\,.
\end{align}
We discarded the $2p^2(yY\xi)$ term which does not contribute at this order.

Taking the time derivative, we find 
\begin{align}
\label{Ceff-dot}
\begin{split}
\dot{C}_\mathrm{eff}|_{q^2 y^2} &=  X_A + q (X_B + X_C +  X_D) \,,
    \\
    X_A &= - 2p YY(\dot{p}|_{q^2y^2}) 
     \\
    X_B &=  2(\xi \dot{\tilde{A}}|_1+ \xi F\dot{y}|_1)(y\tilde{Y} p)  \,,
        \\
     X_C &= 2(\xi \tilde{A}) \left[ (\dot{y}|_1 \tilde{Y} p)  + (y \tilde{Y} \dot{p}|_1) + (y \dot{\tilde{Y}}|_1 p)\right] \,,
    \\
    X_D &= 2y^2(\xi A) (\xi \dot{p}|_1) \,.
\end{split}
\end{align}
A bit of algebra shows that 
\begin{align}
\begin{split}
    m X_A &= \textcolor{blue}{-4q^2pYY(h+\bar{h})} \,,
    \\
    m X_B &= 2q(\xi \tilde{F}\tilde{F}y+ \xi FFy)(y\tilde{Y}p) = \textcolor{purple}{8q (\xi F^+ F^-y) (y\tilde{Y}p)} \,,
\\
      mX_C &= 2q (\xi \tilde{A})\left[ y^2 (\xi \tilde{F}p) \textcolor{brown}{-(\xi y)(y \tilde{F}p)+ (\xi y)  (y\tilde{F} p)} \right] \,,
    \\
    m X_D &= 2q y^2(\xi A) (\xi Fp)\,.
\end{split}
\label{X-ABCD}
\end{align}
A partial cancellation between $X_C$ and $X_D$ leads to 
\begin{align}
    m(X_C+X_D) = 4qy^2 \left[ (\xi A^+) (\xi F^-p)+(\xi A^-) (\xi F^+p)\right] \,.
    \label{X-CD-simple} 
\end{align}
Since we have expanded $C_0$, $C_1$ up to $y^2$ terms, we should try to absorb the whole RHS of \eqref{Ceff-dot} into a $\mathcal{O}(q^2y^2)$ correction to the charge \eqref{Carter-to-quadratic}. 
We propose an ansatz
\begin{align}
\begin{split}
        C_2 &= a_1 f_1    + a_2 f_2 \,,
        \\
        f_1 &= y^2(\xi A^+)(\xi A^-) \,,
        \\
        f_2 &= (\mathcal{R}_-^2+\mathcal{R}_+^2) (yF^+ F^-y) \,.
\end{split}
\label{C2-ansatz} 
\end{align}
The time derivatives of $f_i$ in \eqref{C2-ansatz} are 
(keeping the leading order terms only)
\begin{align}
\begin{split}
      m \dot{f}_1 &= - y^2\left[ (\xi A^+)(\xi F^-p)  + (\xi A^-)(\xi F^+p) \right] \,, 
      \\
      m \dot{f}_2 &= \textcolor{blue}{-4 pYY(h +\bar{h}) }
      \\
      &\quad -2y^2 \left[ (\xi A^+)(\xi F^-p)+ (\xi A^-)(\xi F^+p)\right]
      \\
      &\quad +4 (\xi y) \left[ (\xi A^+)(y F^-p)+ (\xi A^-)(y F^+p)\right]
      \\
      &\quad \textcolor{purple}{-8(\xi \tilde{Y} y) (p F^+ F^- y)  + 8 (\xi \tilde{Y} p) (y F^+ F^- y)}
\end{split}
\end{align}
Note that 
\begin{align}
\begin{split}
   &m\left( X_A + qX_B - q^2 \dot{f}_2 \right) 
   \\
   &\supset 8\left[ (y \tilde{Y} p) (\xi F^+ F^- y) + (\xi \tilde{Y} y) (p F^+ F^- y)  +  (p \tilde{Y} \xi) (y F^+ F^- y) \right]
   \\
   &= 8wYF^+F^-y
   = 8w(Y^++Y^-)F^+F^-y
   \\
   &= 4i \left[ (\xi A^+)(wF^-y)- (\xi A^-)(wF^+y)\right]
   \\
   &= 4 (\xi A^+)[-y^2(\xi F^-p)+(\xi y)(yF^-p)] + \cc \,.
\end{split}
\end{align}
Adding the other terms from $\dot{f}_2$, we get 
\begin{align}
\begin{split}
   m\left( X_A + qX_B - q^2\dot{f}_2 \right) 
  &= -2 q^2 y^2\left[ (\xi A^+)(\xi F^-p) + (\xi A^-)(\xi F^+p)\right] \,.
\end{split}
\end{align}
Adding the contribution from $X_C+X_D$ \eqref{X-CD-simple}, we get 
\begin{align}
\begin{split}
   m\left( X_A + qX_{B+C+D} - q^2\dot{f}_2 \right) 
  &= 2 q^2 y^2\left[ (\xi A^+)(\xi F^-p) + (\xi A^-)(\xi F^+p)\right] \,.
\end{split}
\end{align}
But, this can be canceled by $\dot{f}_1$:
\begin{align}
\begin{split}
   m\left( X_A + qX_{B+C+D}  + q^2(2\dot{f}_1 - \dot{f}_2) \right) 
  &= 0  \,.
\end{split}
\end{align}
Thus we have shown that the desired $\mathcal{O}(q^2y^2)$ correction to $C$ is  
\begin{align}
       C_2 = 2f_1 - f_2
       = 2 y^2(\xi A^+)(\xi A^-) - (\mathcal{R}_-^2+\mathcal{R}_+^2) (yF^+ F^-y) \,.
\end{align}
Finally, using the identities, 
\begin{align}
    (F^\pm Y^\pm)^\m{}_\n = \pm\frac{i}{2} (\xi A^\pm)\delta^\m{}_\n \,,   
    \quad (Y^\pm Y^\pm)_{\m\n} = -\frac{1}{4} \mathcal{R}_\mp^2 \eta_{\m\n} \,,
\end{align}
we obtain a remarkably simple expression for $C_2$: 
\begin{align}
    C_2 = 4yF^+YYF^-y \,.
\end{align}

In summary, the final form of the Carter charge is 
\begin{align}
\begin{split}
    C &= C_0 + q C_1 + q^2 C_2 \,,
    \\
    C_0 &= - pYYp   +2 p^2 (yY \xi) + 2 (\xi p)(y Y p) + p^2 (\xi y)^2  + y^2 (\xi p)^2 \,,
    \\
    C_1 &= 4 [\xi A^+(z)][ y Y^-(z) p] +  4 [\xi A^-(\bar{z})][ y Y^+(\bar{z}) p] \,,
    \\
    C_2 &= 4yF^+YYF^-y \,.
\end{split}
\end{align}
It is conserved to all orders in $y$ at 1PL, and up to the $y^2$ order at 2PL.

%\newpage 
\subsection{Coulomb limit} 

The original Carter charge for a point particle probe 
converges to the angular-momentum-squared $(\vec{L}^2)$ in the Coulomb limit $(a\rightarrow 0)$. 
A similar statement holds for the spinning probe. 
Recall the expression for the free $(q=0)$ Carter charge: 
\begin{align}
\begin{split}
    C_0 &=  (Y_{\nu\rho}  p^\rho)^2
    +2 p^2 (yY \xi) + 2 (\xi p)(y Y p) +w^2  + \xi^2  y^2 p^2 \,.  
\end{split}
\label{Carter-simple-copy}
\end{align}
In the Coulomb limit, it is straightforward to show that 
\begin{align}
    (Y_{\nu\rho}  p^\rho)^2 = \vec{L}^2 \,,
    \quad 
    2 p^2 (yY \xi) = 0 \,,
    \quad 
     2 (\xi p)(y Y p) = 2\vec{L}\cdot \vec{S} \,,
     \quad 
     w^2  + \xi^2  y^2 p^2= \vec{S}^2 \,, 
     \label{Carter-Coulomb}
\end{align}
where $\vec{L}$ and $\vec{S}$ are the orbital and spin angular momenta in the rest frame of the source. 
Hence, we arrive at the expected statement, 
\begin{align}
    C_0|_\mathrm{Coulomb} = \vec{J}^2 = (\vec{L} + \vec{S})^2 \,,
    \label{Carter-J-squared}
\end{align}

In the Coulomb limit, $C$ is not an extra conserved charge. It is rather a combination of Killing momenta. 
As such, it is clear how to carry it over to the interacting theory. 
We take the formula for the Killing momenta \eqref{Killing-momentum-final} 
and replace $K^\m$ with $Y^{\m\n}$ to find 
\begin{align}
    J^\mu = Y^{\mu\nu}(x) p_\nu + y^\mu (\xi p) - p^\mu (\xi y) + q \left[ Y^{\mu\nu}(z) A^+_\nu(z) +  Y^{\mu\nu}(\bar{z}) A^-_\nu(\bar{z})\right]  \,.
\end{align}
It is a 3-vector in disguise since $\xi\cdot J = 0$. 
An interesting feature is that 
when the $q$-linear term is expanded in $y$, all the even order terms vanish, 
reflecting restoration of parity symmetry in the Coulomb limit. 
For example, at $\mathcal{O}(y^0)$ we have $Y^{\m\n} A_\n(x) = 0$, at $\mathcal{O}(y^2)$ we have $y^\a \partial_\a(Y^{\m\r} F_{\r\s})y^\s =0$ and so on. 
The odd order terms in $y$ can be expressed in terms of $(Y\tilde{F}y)$ and its derivatives.

Note that, in the Coulomb limit, the full Carter charge is \emph{exactly} quadratic in $q$:
\begin{align}
\label{C-exact-Coulomb}
\begin{split}
        C = J_\mu J^\mu 
        &= C_0 + qC_1 + q^2 C_2 \,, 
        \\
        C_1 &= 2 \left[ Y_{\mu\nu}(x) p^\nu + y_\mu (\xi p) - p_\mu (\xi y)\right] \left[ Y^{\mu\rho}(z) A^+_\rho(z) +  Y^{\mu\rho}(\bar{z}) A^-_\rho(\bar{z})\right] \,, 
        \\
        C_2 &= \left[ Y_\mu{}^\nu(z) A^+_\nu(z) +  Y_\mu{}^\nu(\bar{z}) A^-_\nu(\bar{z})\right] \left[ Y^{\mu\rho}(z) A^+_\rho(z) +  Y^{\mu\rho}(\bar{z}) A^-_\rho(\bar{z})\right] \,.
\end{split}
\end{align}
But it seems non-trivial to show that this $C_1$ agrees with the $C_1$ \eqref{Carter-q1-final} in the $a\rightarrow 0$ limit.
Let us call $C_1$ in \eqref{C-exact-Coulomb} $C_1$($J^2)$ and the one in \eqref{Carter-q1-final} $C_1$(EoM). 
It is straightforward to expand the two in powers of $y$ and compare them order by order:
\begin{align}
    C_1 = c_{1,0} + c_{1,1}  + c_{1,2}  + \cdots \,.
\end{align}
At the 0th order, we find $c_{1,0}\mbox{(EoM)} = 0 = c_{1,0}(J^2)$. 
At the 1st order, we find 
\begin{align}
    c_{1,1}(J^2) = 2 p Y(x)Y(x)\tilde{F}(x)y = 2 [\xi A(x)][yY(x) p] = c_{1,1}\mbox{(EoM)} \,.
\end{align}
At the 2nd order, we get
\begin{align}
\begin{split}
   c_{1,2}(J^2) &= -2(\xi p)[yY\tilde{F}y]+2(\xi y)[pY\tilde{F}y]  
   \\
   &= 2(\xi A)(\xi p)y^2+2[y\tilde{F}\xi][yYp]-2[yF\xi][y\tilde{Y}p] = c_{1,2}\mbox{(EoM)} \,.
\end{split}
\end{align}
The matching continues to higher order in $y$, but we have not been able to find an all-order-in-$y$ proof. 
Finally, truncating $C_2(J^2)$ at the $y^2$ order, we confirm the expected matching 
at the $q^2 y^2$ order, 
\begin{align}
C_2(\mbox{EoM})|_{y^2} = y \tilde{F} YY\tilde{F} y = C_2(J^2)|_{y^2} \,. 
\end{align}

%\newpage 
%%%%%%%%%%%%%%%%%%%%%%%%%%%%%%%%%%%%%%%
\section{Asymptotic conservation} 
\label{sec:asymptotic}

We turn our attention to the asymptotic conservation in the sense defined in \cite{Akpinar:2025tct}. 
By asymptotic conservation we mean that the asymptotic form of the conserved charges, where all interaction terms can be dropped, commute with the radial action under the Poisson-Dirac algebra. 
Ref.~\cite{Akpinar:2025tct} performed a comprehensive test of the asymptotic conservation covering all PM data available in the literature. In what follows, we will confirm that similar results hold for the root-Kerr source-probe system. 

\subsection{1PL}

\paragraph{Probe lmit} 

Ref.~\cite{Akpinar:2025tct} expresses the conserved charges in terms of asymptotic phase space variables. 
In our conventions, the expressions are
\begin{align}
\begin{split}
     \tilde{E} &= \gamma\,,
    \\
    \tilde{L} &= l \cdot s_2 + \gamma (s_1 \cdot s_2) + (s_1\cdot v_2)(s_2 \cdot v_1) \,,
    \\
    \tilde{R} &= -l \cdot s_1 + \gamma (s_1 \cdot s_2) + (s_1\cdot v_2)(s_2 \cdot v_1) \,,
    \\
    \tilde{C} &= l^2 + 2\gamma\, l\cdot (s_1-s_2) + \gamma^2 (s_1-s_2)^2 - (s_1+s_2)^2 
    \\
    &\qquad  - (s_1\cdot v_2)^2 - (s_2\cdot v_1)^2 -2\gamma (s_1\cdot v_2)(s_2\cdot v_1) \,.
\end{split}
\label{charges-asymptotic}
\end{align}
Here, $s_i$ are the spin-length vectors $(y_i = - s_i)$, and 
\begin{align}
    l_\mu = -\epsilon_{\mu\nu\rho\sigma} (x_1-x_2)^\nu v_1^\rho v_2^\sigma \,.
\end{align}
The tilde in \eqref{charges-asymptotic} denotes rescaling by powers of the probe mass, so that $\tilde{E}$ is dimensionless while $\tilde{L}$, $\tilde{R}$, $\tilde{C}$ all have a length-squared dimension. 
The change of convention from ref.~\cite{Akpinar:2025tct} to this note is as follows:
\begin{align}
    (u\cdot v)_\mathrm{there} = - (u\cdot v)_\mathrm{here} \,,
    \quad 
    l_\mathrm{there} = -  l_\mathrm{here} \,,
    \quad 
    \tilde{R}_\mathrm{there} = - \tilde{R}_\mathrm{here} \,, 
    \quad 
    y_\mathrm{there} = \gamma_\mathrm{here} \,.
\end{align}

Let us check whether \eqref{charges-asymptotic} agrees with the free theory limit of the Noether charges we used in other sections. The energy is trivial. The angular momentum $L$ of the probe is 
\begin{align}
    L = x^1 p^2 - x^2 p^1 + y^0 p^3- y^3 p^0 \,. 
\end{align}
With $v_2 = \xi = (1,0,0,0)$ and $s_2 = (0,0,0,a)$, we find 
\begin{align}
    \tilde{L} = \frac{a}{m_1} L \,.
\end{align}
As for the R\"udiger charge, we note that 
\begin{align}
   R =  yY(x)p = \vec{y} \cdot(\vec{x}\times \vec{p}) + a (y^0 p^3- y^3 p^0) \,.
\end{align}
It agrees with \eqref{charges-asymptotic} as $\tilde{R} = R/m_1$ since 
\begin{align}
    -l \cdot s_1 = \frac{1}{m_1} \vec{y} \cdot(\vec{x}\times \vec{p}) \,,
    \quad 
    \gamma (s_1 \cdot s_2) + (s_1\cdot v_2)(s_2 \cdot v_1) = \frac{a}{m_1} (y^0 p^3- y^3 p^0) \,.
\end{align}
Finally, recall that the Carter charge is 
\begin{align}
       C &= - pYYp + 2p^2 (yY\xi) + 2(\xi p)(y Yp) +p^2(\xi y)^2 + y^2(\xi p)^2 \,.
\end{align}
Comparing it with \eqref{charges-asymptotic}, we find an agreement up to a constant shift, 
\begin{align}
    \tilde{C} + (s_1^2+s_2^2) = \frac{1}{m^2} C \,.
\end{align}

\paragraph{Arbitrary mass ratio} 

In eqs.(37)-(39) of ref.~\cite{Akpinar:2025tct}, the asymptotic charges for arbitrary mass ratio are given.
In our conventions, 
\begin{align}
    \begin{split}
        \mathcal{L} &= \frac{1-\rho}{2} (l\cdot s_1) + \frac{1+\rho}{2} (l\cdot s_2)  + (\textcolor{blue}{\nu} + \gamma)(s_1\cdot s_2) + (s_1\cdot v_2)(s_2 \cdot v_1) \,,
        \\
        \mathcal{R} &= - \frac{1+\rho}{2} (l\cdot s_1) - \frac{1-\rho}{2} (l\cdot s_2)  + (\textcolor{blue}{\nu} + \gamma)(s_1\cdot s_2) + (s_1\cdot v_2)(s_2 \cdot v_1) \,,
        \\
        \mathcal{C} &= l^2 + 2\textcolor{blue}{\rho}\, \gamma\, l\cdot (s_1-s_2) + \gamma^2 (s_1-s_2)^2 - (s_1+s_2)^2 
        \\
        &\qquad  - (s_1\cdot v_2)^2 - (s_2\cdot v_1)^2 -2\gamma (s_1\cdot v_2)(s_2\cdot v_1) - 4\textcolor{blue}{\nu}\,\gamma (s_1\cdot s_2)\,.
    \end{split}
    \label{charges-asymptotic-1PM}
\end{align}
We introduced shorthand notations for dimensionless ratios
\begin{align}
    \nu = \frac{2m_1m_2}{m_1^2 +m_2^2} \,,
    \quad 
    \rho = \frac{-m_1^2+m_2^2}{m_1^2+m_2^2} \,.
\end{align}
Together with the energy $\mathcal{E} = \gamma$, the four charges all Poisson-commute with each other. 
In the probe limit $(m_1/m_2 \rightarrow 0)$, the expressions reduce to \eqref{charges-asymptotic} as expected.

It may be useful to work with the linear combinations, 
\begin{align}
\begin{split}
     \mathcal{L} - \mathcal{R} &= l\cdot (s_1 + s_2)  
     \\
      \mathcal{L} + \mathcal{R} &= -\rho\,l\cdot (s_1 - s_2)  + 2(\nu+\gamma)(s_1\cdot s_2) + 2 (s_1\cdot v_2)(s_2 \cdot v_1) \,.
\end{split}
\end{align}

\paragraph{Radial action}

The radial action can be written in terms of the asymptotic charges 
\cite{Akpinar:2025tct}. 
It is well known that the 1PL radial action $\chi_\mathrm{1PL}$ depends only on $y = y_1 + y_2 = -(s_1 +s_2)$. Then $\mathcal{C}$ and $\mathcal{L} + \mathcal{R}$ cannot independently contribute to $\chi_\mathrm{1PL}$. 
The suitable combination is 
\begin{align}
\begin{split}
     \mathcal{C}+ 2\gamma(\mathcal{L} + \mathcal{R}) 
     &= l^2 + (\gamma^2-1) s^2  - (s\cdot v_2)^2 - (s\cdot v_1)^2 + 2\gamma (s \cdot v_2) (s \cdot v_1) 
     \\
     &= (\gamma^2-1)(b^2 + y_\perp^2) \,.
\end{split}
\end{align}
Here, we used $l^2 = b^2( \gamma^2-1)$, $s_a \cdot v_a = 0$, and 
\begin{align}
\begin{split}
      (\epsilon_\mu[y, v_1, v_2])^2 &= y_\perp^2 (\gamma^2-1) 
      \\
      &= (\gamma^2-1) y^2  - (y\cdot v_2)^2 - (y\cdot v_1)^2 + 2\gamma (y \cdot v_2) (y \cdot v_1) \,. 
\end{split}
\end{align}
We also note that 
\begin{align}
 \frac{\mathcal{L} -\mathcal{R}}{\sqrt{\gamma^2-1}} = - \frac{\epsilon[b,y,v_1,v_2]}{\sqrt{\gamma^2-1}}  \,,
 \qquad \left| \frac{\epsilon[b,y,v_1,v_2]}{\sqrt{\gamma^2-1}}\right| =
    \sqrt{b^2 y_\perp^2 - (b\cdot y_\perp)^2} \,.
\end{align}
The final result for $\chi_\mathrm{1PL}$ is \cite{Kim:2024grz} 
\begin{align}
\begin{split}
    \chi_\mathrm{1PL} &= \frac{q_1q_2}{8\pi} \sum_{\sigma = \pm 1} \left( \frac{\gamma}{\sqrt{\gamma^2 -1}} +\sigma \right) \log\left(\frac{\mathcal{C}+ 2\gamma(\mathcal{L} + \mathcal{R}) }{\gamma^2-1}  + \sigma \frac{2(\mathcal{L}-\mathcal{R})}{\sqrt{\gamma^2-1}}\right) 
    \\
    &=\frac{q_1q_2}{8\pi\sqrt{\gamma^2-1}} \sum_{\sigma = \pm 1} e^{\sigma \theta}\log\left(\frac{\mathcal{C}+ 2( e^{\sigma\theta}\mathcal{L} +  e^{-\sigma\theta}\mathcal{R}) }{\gamma^2-1} \right)  \,, 
    \quad 
    \gamma = \cosh\theta \,,\; \theta > 0 \,.
\end{split} 
\end{align}
In the same notation, the 1PM classical eikonal for a pair of Kerr black holes is \cite{Akpinar:2025tct}
\begin{align}
   \chi_\mathrm{1PM} &= - \frac{Gm_1 m_2}{2\sqrt{\gamma^2-1}} \sum_{\sigma = \pm 1} e^{2\sigma \theta}\log\left(\frac{\mathcal{C}+ 2( e^{\sigma\theta}\mathcal{L} +  e^{-\sigma\theta}\mathcal{R}) }{\gamma^2-1} \right) \,.
\end{align}

%\newpage
\subsection{2PL} 

In this subsection, we check the asymptotic conservation at 2PL, order by order in spin, 
using the classical eikonal computed in \cite{Kim:2024grz}. 
We take the probe limit $(m_2\rightarrow \infty)$ of the 2PL eikonal and expand the result in powers of spin: $y_1^{k_1} y_2^{k_2}$ up to $k_1 + k_2 \leq 3$. We remove an overall coefficient by defining
\begin{align}
    \chi_\mathrm{2PL}(m_2 \rightarrow \infty) = \frac{(q_1 q_2)^2}{32\pi m_1(\gamma^2-1)^{1/2}} F(b,y_1, y_2,v_1,v_2) \,.
\end{align}

Before we expand $F$ in powers of spin, 
we prepare some building blocks for the algebra that we need to establish charge conservation. 
We discuss the R\"udiger charge first and proceed to the Carter charge. 
Since $F$ carries some powers of $|b|$, we need 
\begin{align}
        \{ R, b^2 \} &= -2 (b\cdot y_1) (y_1\cdot v_2)  -2 (b\cdot y_2) (y_1\cdot v_2) \,. 
\end{align}
The dot products appear in the numerators of $F$. So, we also need
\begin{align}
    \begin{split}
        \{R, b\cdot y_1 \} &= (y_1\cdot v_2) (b^2 - y_1^2 -y_1\cdot y_2) + \epsilon[b,v_2,y_1,y_2] 
        \\
        &\qquad + \frac{1}{\gamma^2-1}\left[ (y_1\cdot v_2)^3 
        - \gamma(y_1\cdot v_2)^2(y_2\cdot v_1) \right] \,,
        \\
        \{R, b\cdot y_2 \} &= -(y_1\cdot v_2)(y_1\cdot y_2+y_2^2) - \epsilon[b,v_2,y_1,y_2]
        \\
        &\qquad  + \frac{(y_1\cdot v_2)(y_2 \cdot v_1)}{\gamma^2-1}\left[ (y_2\cdot v_1) - \gamma (y_1\cdot v_2)\right]\,,
           \\
        \{R, y_1 \cdot v_2 \} &= -(\gamma^2-1) (b\cdot y_1) - \gamma\, \epsilon[v_1, v_2, y_1, y_2] \,,
           \\
        \{R, y_2 \cdot v_1  \} &= - \epsilon[v_1, v_2, y_1, y_2] \,.
    \end{split}
\end{align}
For the epsilon products on the RHS, we have
\begin{align}
\begin{split}
        \{ R, \epsilon[b,v_1,v_2,y_1] \} &= (y_1\cdot v_2)\epsilon[v_1,v_2,y_1,y_2] - (b\cdot y_1)(y_2\cdot v_1) -\gamma (b\cdot y_2)(y_1\cdot v_2)  
        \\
          &= - \{ R, \epsilon[b,v_1,v_2,y_2] \} \,,
          \\
         \{ R, \epsilon[v_1,v_2,y_1,y_2] \} &= (y_2\cdot v_1)y_1^2+(y_1\cdot v_2)^2(y_2\cdot v_1) + \gamma (y_1\cdot v_2)(y_1\cdot y_2) 
         \\
         &\quad + (y_1\cdot v_2) \epsilon[b,v_1,v_2,y_2]  + (y_2\cdot v_1) (y_1\cdot y_2) +\gamma (y_1\cdot v_2) y_2^2 \,.
\end{split}
\end{align}
For the bi-linears of $y_{1,2}$, we have 
\begin{align}
    \begin{split}
        \{ R, y_1^2 \} &= 0 \,,
        \\
        \{ R, y_2^2 \} &= 0 \,,
        \\
        \{ R, y_1\cdot y_2 \} &= (b\cdot y_2)(y_1\cdot v_2) + \gamma (b\cdot y_1) (y_2\cdot v_1) +(y_2\cdot v_1) \epsilon[v_1,v_2,y_1,y_2] \,.
    \end{split}
\end{align}
We have five vectors $(b,v_1, v_2, y_1, y_2)$. Out of the five $\epsilon$ products, only three are independent, 
since we can use the Schouten identity to show that 
\begin{align}
\begin{split}
    \gamma\, \epsilon[b,v_1,y_1,y_2] - \epsilon[b,v_2,y_1,y_2] &= (v_1\cdot y_2) \epsilon[b,v_1,v_2,y_1] \,,
    \\
    -\epsilon[b,v_1,y_1,y_2] + \gamma\, \epsilon[b,v_2,y_1,y_2] &= (v_2\cdot y_1) \epsilon[b,v_1,v_2,y_2] \,.
\end{split}
\end{align}
The dot products and the $\epsilon$ products form a complete set of building blocks.

Now, we list the expansion of $F$ in powers of $y_{1,2}$. 
The spinless part is 
\begin{align}
    \begin{split}
    F_0 \equiv F|_{y^0} = \frac{1}{|b|} \,.
    \end{split}
\end{align}
The linear in spin term is 
\begin{align}
   F|_{y^1} = \frac{\gamma  }{(\gamma^2-1) |b|^3}  
   \left( \epsilon[b,v_1,v_2,y_1]+2 \epsilon[b,v_1,v_2,y_2] \right) \,.
\end{align}
The quadratic term is 
\begin{align}
F|_{y^2} =
\frac{1}{4(\gamma^2-1)^2 |b|^{3}}
(\mathcal{K}_\mathrm{NJ} + \mathcal{K}_\mathrm{contact} ) \,, 
\end{align}
where 
\begin{align}
    \begin{split}
     \mathcal{K}_\mathrm{NJ} =&(3\gamma^4-11\gamma^2+2)(v_2\cdot y_1)^2+(-13\gamma^2+7)(v_1\cdot y_2)^2\\&\quad\quad\quad+4(4\gamma^2-1)(v_2\cdot y_1)(v_1 \cdot y_2)\\&+2(\gamma^2-1)\left[(3\gamma^2-1)y_1^2+(5\gamma^2-3)y_2^2+(8\gamma^2-4)(y_1\cdot y_2)\right]\\&
     -\frac{3}{b^2}[(3\gamma^4-4\gamma^2+1)(b\cdot y_1)^2+(5\gamma^4-8\gamma^2+3)(b\cdot y_2)^2\\&\quad\quad\quad+4(2\gamma^4-3\gamma^2+1)(b\cdot y_1)(b\cdot y_2)] \,,
    \end{split}
\end{align}
and 
\begin{align}
    \begin{split}
     \mathcal{K}_\mathrm{contact} = (\gamma^2-1)\left[(4-3\gamma^2)(v_2\cdot y_1)^2-2(\gamma^2-1)y_1^2+\frac{3}{b^2}(\gamma^2-1)(b\cdot y_1)^2\right] \,.
    \end{split}
\end{align}
Adding the non-contact and contact terms, we obtain the final form of $\mathcal{K}$, 
\begin{align}
\begin{split}
       \mathcal{K}_\mathrm{all} &= A_{11} (v_2\cdot y_1)^2 + A_{22} (v_1\cdot y_2)^2 + A_{12} (v_2\cdot y_1)(v_1\cdot y_2) 
       \\
       &\quad + 2(\gamma^2-1)\left[ C_{11} y_1^2 + C_{22} y_2^2 + C_{12} (y_1\cdot y_2)\right] 
       \\
       &\quad - \frac{3}{b^2}\left[ D_{11} (b\cdot y_1)^2 + D_{22}(b\cdot y_2)^2 +D_{12} (b\cdot y_1)(b\cdot y_2) \right] \,,
\end{split}
\end{align}
with the coefficients 
\begin{align}
    \begin{split}
        A_{11} = -4\gamma^2-2 \,,\quad &A_{22} = -13\gamma^2+7 \,, \quad A_{12} =  4\gamma(4\gamma^2-1)\,,
        \\
        C_{11} = 2\gamma^2 \,, \quad &C_{22} = 5\gamma^2-3 \,, \qquad C_{12} = 4 (2\gamma^2-1) \,,
        \\
        D_{11} = 2\gamma^2(\gamma^2-1)  \,, \quad &D_{22} = 5\gamma^4-8\gamma^2+3 \,,\quad D_{12} = 4 (2\gamma^4-3\gamma^2+1) \,.
    \end{split}
\end{align}
Similarly, the cubic term is given by 
\begin{align}
    F|_{y^3} = \frac{3}{4(\gamma^2-1)^2|b|^5} (\mathcal{H}_\mathrm{NJ} + \mathcal{H}_\mathrm{contact} ) \,,
\end{align}
where 
\begin{align}
    \begin{split}
     \mathcal{H}_\mathrm{NJ} =&2\gamma(\gamma^2-1)\left[(b\cdot y) \epsilon [ v_1,v_2,y_1,y_2 ]+y^2[2 \epsilon [ b,v_1,v_2,y_2 ]+ \epsilon [ b,v_1,v_2,y_1 ]]\right]\\
     &+2[\gamma(\gamma^3-3)(v_2\cdot y_1)^2 -3\gamma(v_1\cdot y_2)^2 
     \\&+(-\gamma^3+5\gamma^2+1)(v_2\cdot y_1)(v_1 \cdot y_2)] \epsilon [ b,v_1,v_2,y_2 ]\\
     &+[\gamma(\gamma^2-4)(v_2\cdot y_1)^2+\gamma(2\gamma^2-2\gamma-5)(v_1\cdot y_2)^2\\&+2(-\gamma^4+\gamma^3+3\gamma^2+1)(v_2\cdot y_1)(v_1\cdot y_2)] \epsilon [ b,v_1,v_2,y_1 ]
     \\&-\frac{5}{b^2}\gamma(\gamma^2-1)(b\cdot y)^2[2 \epsilon [ b,v_1,v_2,y_2 ]+ \epsilon [ b,v_1,v_2,y_1 ]] \,,
    \end{split}
\end{align}
and 
\begin{align}
    \begin{split}
     \mathcal{H}_\mathrm{contact} = -2(\gamma^2-1)\left[(v_2\cdot y_1)^2 \epsilon [ b,v_1,v_2,y_2 ]+(v_2\cdot y_1)(v_1\cdot y_2) \epsilon [ b,v_1,v_2,y_1 ] \right] \,. 
    \end{split}
\end{align}
Adding the two terms, we find 
\begin{align}
    \begin{split}
        \mathcal{H}_\mathrm{all} = &2\gamma(\gamma^2-1)\left[(b\cdot y) \epsilon [ v_1,v_2,y_1,y_2 ]+y^2[2 \epsilon [ b,v_1,v_2,y_2 ]+ \epsilon [ b,v_1,v_2,y_1 ]]\right]\\
     &+2[{E}_{11}(v_2\cdot y_1)^2 +{E}_{22}(v_1\cdot y_2)^2 
     +{E}_{12}(v_2\cdot y_1)(v_1 \cdot y_2)] \epsilon [ b,v_1,v_2,y_2 ]\\
     &+[\tilde{E}_{11}(v_2\cdot y_1)^2+\tilde{E}_{22}(v_1\cdot y_2)^2+\tilde{E}_{12}(v_2\cdot y_1)(v_1\cdot y_2)] \epsilon [ b,v_1,v_2,y_1 ]
     \\&-\frac{5}{b^2}\gamma(\gamma^2-1)(b\cdot y)^2[2 \epsilon [ b,v_1,v_2,y_2 ]+ \epsilon [ b,v_1,v_2,y_1 ]] \,,
    \end{split}
\end{align}
with the coefficients
\begin{align}
    \begin{split}
        &E_{11} = \gamma^4-\gamma^2-3\gamma+1 \,,\quad E_{22} = -3\gamma \,, \quad E_{12} =  -\gamma^3+5\gamma+1\,,
        \\
        &\tilde{E}_{11} = \gamma(\gamma^2-4) \,, \quad \tilde{E}_{22} = \gamma(2\gamma^2-2\gamma-5) \,, \quad \tilde{E}_{12} = 2(-\gamma^4+3\gamma^2+\gamma+1) \,.
    \end{split}
\end{align}
A computation shows that $\{C,F+F_{\text{contact}}\}$ vanishes up to $y^3$. 
In contrast, 
we obtain a simple non-zero contribution from $\{R,F+F_{\text{contact}}\}$ 
\begin{align}
    \{R,F+F_{\text{contact}}\}\vert_{y^3} = -\frac{3 \gamma(b \cdot y_1)
   (v_2 \cdot y_1) \epsilon[b,v_1,v_2,y_1]}{2 b^5}\,.
   \label{Rudiger-3rd-nonzero}
\end{align}

Our final question is whether we can add a new $y_1^3$ term to $F$, attributable to an additional contact term, to restore conservation. Specifically, we work with an ansatz, 
\begin{align}
    F_\mathrm{new} = \frac{N_\mathrm{new}}{|b|^5} \,, 
    \qquad 
    N_\mathrm{new} \propto b \, y_1^3 \,,
\end{align}
such that 
\begin{align}
    \{ C , F_\mathrm{new} \}|_{y^3} = 0 \,,
    \quad 
    \{ R , F_\mathrm{new} \}|_{y^3} = +\frac{3 \gamma(b \cdot y_1)
   (v_2 \cdot y_1)\epsilon[b,v_1,v_2,y_1]}{2 b^5}\,.
\end{align}
The most general possibility for $N_\text{new}$ is 
\begin{align}
\label{N-new-ansatz}
    N_\mathrm{new} = \left(\alpha_1 y_1^2+\alpha_2 (v_2\cdot y_1)^2 + \alpha_3 \frac{(b\cdot y_1)^2}{b^2}\right) \epsilon[b,v_1,v_2,y_1] \,.
\end{align}
The coefficients $\alpha_{1,2,3}$ do not affect the conservation of the Carter charge up to the $y^3$ order. The restoration of the conservation of the R\"udiger charge imposes one constraint on $\alpha_{2,3}$ 
and no constraint on $\alpha_1$. 

It remains to see if the desired linear combination of the $\alpha_2$ and $\alpha_3$ terms can be produced by a new interaction vertex of our world-line model. 
The exhaustive list of the interaction vertices, in terms of Hamiltonian deformation, are 
\begin{align}
H_2' = \frac{1}{p^2} \left[ \beta_1 G_1 + \beta_2 G_2 + \beta_3 G_3 + \beta_4 G_4 \right] \,.
\end{align}
The functions $G_i$ ($i=1,2,3,4$) are 
\begin{align}
    \begin{split}
      G_1 &= i y^\mu \left[ (y \partial_\mu F^+ p) ( yF^-p) - (y F^+p) (y \partial_\mu  F^- p) \right] \,,
      \\
       G_2  &= ip^2\, y^\mu \left( y \partial_\mu F^+ F^-y - yF^+ \partial_\mu F^- y\right)\,,
        \\
       G_3 &= i y^2y^\mu \left(p \partial_\mu F^+ F^- p - p F^+ \partial_\mu F^-p \right) \,,
        \\
       G_4 &= i y^2 p^\mu \left( y \partial_\mu F^+ F^- p - y F^+ \partial_\mu F^-p  \right)\,.
    \end{split}
    \label{3rd-integrand}
\end{align}
Here, all $F^\pm$, $\partial_\m F^\pm$ are evaluated at $x^\mu$. 
The contribution of a vertex factor to the classical eikonal can be computed using the elementary formula \cite{Kim:2024grz,Kim:2025olv}:
\begin{align}
    \chi_\text{contact} = - \int H_\text{contact} d\tau \,. 
\end{align}
The field configuration to enter the vertex factors \eqref{3rd-integrand} is 
\begin{align}
    F^\pm_{\mu\nu}(x)
    = -\frac{q_2}{8\pi \vert\vec{x}\vert^2}
    \left((x\wedge v_2)_{\mu\nu}\mp i\,\epsilon_{\mu\nu}[x,v_2]\right)\,,
\end{align}
where
\begin{align}
    x = b +  v_1 \tau \,,\quad 
    \vert\vec{x}\vert^2 = x^2 + (x\cdot v_2)^2\,.
\end{align}
In the end, all four vertex factors \eqref{3rd-integrand} produce no $\alpha_{2,3}$ type of terms in \eqref{N-new-ansatz} while producing some  $\alpha_1$ type terms, which do not affect the conservation of the R\"udiger charge. 
We conclude that it is impossible to restore the conservation of the R\"udiger charge under the assumptions of this paper. 

%\newpage 
\section{Discussion}
\label{sec:discussion}

For a root-Kerr source-probe system, our analysis shows that 
the Carter and R\"udiger charges 
are conserved to all orders in spin at 1PL, and up to the spin-squared order at 2PL.
under the assumptions of this paper.
The failure of conservation at the spin-cubic order, 
and a parallel problem at the 2PM $S^5$ order in asymptotic conservation \cite{Akpinar:2025tct}, 
are reminiscent of the observation in ref.~\cite{Arkani-Hamed:2017jhn}; 
based on the structure of the Compton amplitudes, it was shown 
that charged particles with (quantum) spin $s\ge 3/2$ and any gravitating particle with spin $s\ge 5/2$ cannot be elementary. 
The analysis of this paper was purely classical and made no direct contact with quantum amplitudes, 
but it would be interesting to explore possible connections following, {\it e.g.}, refs.~\cite{Guevara:2018wpp,Chung:2018kqs,Guevara:2019fsj,Chung:2019duq}. 

Even an approximate conservation can be useful in solving the EoM, either analytically or numerically \cite{Witzany:2019nml,Drummond:2023loz,Skoupy:2024uan,Skoupy:2025nie,Piovano:2025aro}. 
On the analytic side, the radial action (or classical eikonal) depends on the conserved charges. 
For a Kerr black hole background, the radial action was computed for orbits in the equatorial plane in \cite{Damgaard:2022jem}, 
and for generic orbits in \cite{Gonzo:2024zxo}. 
In a parallel way, the explicit form of the Carter and R\"udiger charges in this paper, including the perturbative corrections, could be used to compute the radial action of the root-Kerr probe. 

One notable remaining question is whether the asymptotic conservation up to the $\mathcal{O}(G^2 S^4)$ order observed in ref.~\cite{Akpinar:2025tct} would lead to the stronger notion of local conservation. 
At 1PM, the test of conservation would require an all order in spin world-line action based on the non-linear NJ shift constructed in refs.~\cite{Kim:2025xka,Kim:2026opo,Kim:2026yqo}. 
At 2PM, the enumeration of all relevant curvature-squared and/or higher derivative terms is likely to pose a challenge. 

In the self-dual sector, recent works \cite{Kim:2025xka,Kim:2026opo,Kim:2026yqo} found exact conservation of both R\"udiger and Carter to all orders in
spin. It will be interesting to understand how exactly the ASD part of the root-Kerr field profile (an ASD dyon \cite{Kim:2024dxo,Kim:2024mpy}) spoils the exact conservation of the SD sector to reproduce the conclusion of
this paper.

\acknowledgments

We thank Hojin Lee for collaboration at an early stage of this work. 
We are grateful to Joon-Hwi Kim and Jung-Wook Kim for a series of helpful correspondences, 
and to David Kosower, Sungjay Lee and Piljin Yi for enlightening discussions. 
This work is supported by National Research Foundation of Korea (NRF) grants, NRF-2023-K2A9A1A0609593811 and NRF RS-2024-00351197, as well as KIAS grant PG006002. 

\newpage
\appendix 
\section{Comparison} \label{sec:dFV}

%The matching should proceed with (1) EoM, (2) Killing, (3) R\"udiger, (4) Carter. 

For the reader's convenience, 
we provide a detailed comparison between key results of this paper and those of ref.~\cite{deFirmian:2026mln}. 
The EoM and the conserved charges are matched explicitly up to the $(q^1 y^2)$ order. 

Expanding our EoM to $\mathcal{O}(q^1 y^2)$, we get 
\begin{subequations}
\label{our-EoM}
    \begin{align}
        \dot{p}_\m &= \frac{q}{m} \left(F_{\m\n} + y^\r \partial_\r  \tilde{F}_{\m\n} - \frac{1}{2} y^\r y^\s \partial_\r \partial_\s F_{\m\n} \right)p^\n \,,
        \\
        \dot{y}^\m &= \frac{q}{m} \left(F^{\m\n} + y^\r \partial_\r  \tilde{F}^{\m\n}  \right) y_\n  \,,
        \\
        \dot{x}^\m &= \frac{p^\m}{m} - \frac{q}{m}\left( \tilde{F}^\m{}_\n - y^\r\partial_\r F^\m{}_\n\right) y^\n \,. 
    \end{align}
\end{subequations}
The EoM of ref.~\cite{deFirmian:2026mln} truncated at $\mathcal{O}(q^1 s^2)$ are
\begin{subequations}
\label{dFV-EoM}
\begin{align}
    \frac{d\tilde{p}_\m}{d\tilde{\tau}}  &=  q F_{\m\n} u^\n -\frac{q}{2m} \left( 2 s^\r u^\s \partial_\m \tilde{F}_{\r\s} + s^\n s^\r u^\s \partial_\m \partial_\n F_{\r\s} \right)\,,
    \\
    \label{dFV-s-EoM}
    %\begin{split}
          \frac{d s^\m}{d\tilde{\tau}} &= \frac{q}{m} F^{\m\n} s_\n 
           %\\
           %&\quad 
           + \frac{q}{2m^2}\left[ s^\m u^\n u^\r(\partial_\n \tilde{F}_{\r\s})s^\s + s^2 u^\r (\partial_\r \tilde{F}^\m{}_\n) u^\n -2 s^\r (\partial_\r \tilde{F}^\m{}_\n) s^\n \right] \,, 
    %\end{split}
    \\
    \begin{split}
       \frac{d\tilde{x}^\m}{d\tilde{\tau}} 
       %&= u^\m - \frac{1}{m} \mathcal{N}^{\mu \nu} u_\nu - \frac{1}{m^2} S^{\m\n} \left( \mathcal{F}_\nu+q F_{\nu \rho} u^\rho \right) + \mathcal{O}(q^2, s^3) 
        %\\
        &= u^\m - \frac{q}{2m^3}\left[ u^\m(s^\r s^\s \partial_\r F_{\s\n} u^\n) + s^\r s^\s (\partial_\r F_\s{}^\m)\right]
        \\
        &\qquad \qquad - \frac{3q}{2m^3} s^\m ( u^\r u^\s \partial_\r F_{\s\n} s^\n) +\frac{q}{m^3} s^2 (u^\r u^\s \partial_\r F_\s{}^\m)\,. 
    \end{split}
\end{align}
\end{subequations}

%The ``dynamical mass" is given in Eq.(5.1) of dFV: 
%\begin{align}
%    \mathcal{M}^2 = m^2 + 2q s^\m u^\n \tilde{F}_{\m\n}  + \frac{q}{m} s^\m s^\n u^\r \partial_\m F_{\n\r} \,. 
%\end{align}
%The $s$-EoM comes from Eq.(2.17) of dFV: 
%\begin{align}
%    \dot{s}^\mu=\frac{\alpha}{2}\left(\epsilon^{\mu \nu}{ }_{\rho \sigma} u^\rho s^\sigma \frac{\partial}{\partial s^\nu}-\frac{u^\mu}{\mathcal{M}} s^\nu \frac{\partial}{\partial z^\nu}\right) \mathcal{M}^2+\frac{u^\mu}{\mathcal{M}} q F_{\nu \rho} s^\nu \dot{z}^\rho \,.
%\end{align}
%In \eqref{dFV-s-EoM}, we separated the $s^1$ terms from the $s^2$ terms. 
%
%
%\paragraph{Matching} 
%Matching the Killing charges suggests a redefinition of the momentum:
The variables of the two papers are mapped to each other as follows, 
\begin{subequations}
\label{mapping-dFV-all}
 \begin{align}
    \tilde{p}_\m &= p_\m - q \tilde{F}_{\m\n} y^\n - \frac{q}{2} y^\r y^\s  \partial_\r F_{\s\m} \,, 
    \\
    s^\m &= - m y^\m + \frac{q}{2m}\left( - y^\m ( y \tilde{F} p)  + y^2 \tilde{F}^{\m\n} p_\n \right)  \,, 
    \\
    \tilde{x}^\m &= x^\m + \frac{q}{m^2}\left( \frac{3}{2} y^\m (yFp) - y^2 F^{\m\n} p_\n\right) \,, 
    \label{x-shift-dFV}
\end{align}   
\end{subequations}
The world-line time parameters are related by 
\begin{align}
\label{tau-tilde-vs-tau}
\frac{d\tilde{\tau}}{d\tau} = 1 - \frac{q}{m^2} (y \tilde{F} p)  +  \frac{q}{m^2} (y^\r y^\s \partial_\r F_{\s\n} p^\n) \,.
\end{align}
The velocity vector of ref.~\cite{deFirmian:2026mln} is defined as 
\begin{align}
\begin{split}
      u_\m = \frac{\tilde{p}_\m}{\sqrt{-\tilde{p}^2}} &= \frac{1}{m} \left( p_\m - q \tilde{F}_{\m\n} y^\n - \frac{q}{2} y^\r y^\s  \partial_\r F_{\s\m}\right) 
    \\
    &\qquad + \frac{p_\m}{m^3}  \left( q (y\tilde{F}p) - \frac{q}{2} y^\r y^\s  \partial_\r F_{\s\n} p^\n\right) \,.
\end{split}
\end{align}
It is easy to show that the change of variables \eqref{mapping-dFV-all} and \eqref{tau-tilde-vs-tau} transform \eqref{our-EoM} to \eqref{dFV-EoM} 
and vice versa. 

Proceeding to the conserved charges, we collect all our charges in one place:
\begin{subequations}
    \begin{align}
        Q &= K^\mu(x) p_\mu + \frac{1}{2} \epsilon^{\mu\nu\rho\sigma} y_\mu p_\nu \partial_\rho K_\sigma+ q \left[ K^\mu(z)A_\mu^+(z) +  K^\mu(\bar{z})A_\mu^-(\bar{z}) \right] \,.
\\
R&= y^\m Y_{\m\n} p^\n +  \frac{2q^2}{p^2} \left[ yY(x)p - y^2(\xi p) \right]  [y F^+(z)F^-(\bar{z})y] \,,
\\
\begin{split}
   C&= - pYYp   +2 p^2 (yY \xi) + 2 (\xi p)(y Y p) + p^2 (\xi y)^2  + y^2 (\xi p)^2 \,,
    \\
    &\qquad + 4q [\xi A^+(z)][ y Y^-(z) p] +  4q [\xi A^-(\bar{z})][ y Y^+(\bar{z}) p] + 4q^2 yF^+YYF^-y \,. 
\end{split}
    \end{align}
\end{subequations}
For the comparison, it suffices to truncate the charges at the $y^2$ order, 
\begin{subequations}
\label{our-charges}
    \begin{align}
    \begin{split}
         Q_\text{tr} &= K^\mu (p_\mu + q A_\m)+ \frac{1}{2} \epsilon^{\mu\nu\rho\sigma} y_\mu p_\nu \partial_\rho K_\sigma 
         \\
         &\qquad 
        - q \tilde{F}_{\m\n} K^\m y^\n -\frac{q}{2} y^\r y^\s (\partial_\r K^\n)  F_{\s\n} - \frac{q}{2} K^\m y^\r y^\s  \partial_\r F_{\s\m} \,, 
    \end{split}     
\\
R_\text{tr} &= y^\m Y_{\m\n} p^\n  \,,
\\
\begin{split}
   C_\text{tr} &= - pYYp   +2 p^2 (yY \xi) + 2 (\xi p)(y Y p) + p^2 (\xi y)^2  + y^2 (\xi p)^2 
    \\
    &\qquad + 2q (\xi A)[ y Y p+y^2(\xi p)]
    \\&\qquad + q\left[- 2(\xi \tilde{F}y)(yYp) + 2(\xi Fy)(y\tilde{Y}p) + y^2(\xi p) F^{\r\s} \tilde{Y}_{\r\s}  \right]  \,. 
\end{split}
    \end{align}
\end{subequations}
The charges of ref.~\cite{deFirmian:2026mln} are 
\begin{subequations}
\label{dFV-charges}
    \begin{align}
    \begin{split}
         Q_\text{dFV} &= K^\mu(\tilde{x}) (\tilde{p}_\mu +q A_\m(\tilde{x}))+ \frac{1}{2} \epsilon^{\mu\nu\rho\sigma} u_\mu s_\nu \partial_\rho K_\sigma\,, 
    \end{split}     
\\
R_\text{dFV} &=  u^\m Y_{\m\n} s^\n 
- \frac{q}{2m^2}\left[ 3 (uFs) (u \tilde{Y} s)
-  (s F  \tilde{Y} s)
+2 s^2 (u F \tilde{Y} u) \right]\,,
\\
\begin{split}
   C_\text{dFV} &= \left[- \tilde{p}YY\tilde{p} +2 (\tilde{p} u) (\xi Y s) + 2 (\xi \tilde{p})(u Y s) - (\xi s)^2  + s^2 (\xi u)^2 \right]_{\tilde{x}} 
   \\   
   &\quad  - q (u \tilde{F} s) Y^{\r\s} Y_{\r\s} - q (u \tilde{F} Y Y s) -q (u Y Y \tilde{F} s) +\frac{q}{m} s^\mu s^\nu ( u \tilde{Y} \tilde{Y})^\r \partial_\m F_{\n \r}
   \\
   &\qquad -\frac{q}{m}\left[  s^2  (\xi u) F^{\rho \sigma } \tilde{Y}_{\rho \sigma} -2 (\xi s) (u F \tilde{Y} s)
   +3 (\xi u) (s F \tilde{Y} s) + s^2 (\xi F \tilde{Y} u) \right] \,. 
\end{split}
    \end{align}
\end{subequations}
We flipped the overall sign of the R\"udiger charge: $(R_\text{dFV})_\text{here} = -Q_\text{there}$. 

A straightforward but tedious computation using the map \eqref{mapping-dFV-all} shows that 
our charges in 
\eqref{our-charges} and the charge of ref.~\cite{deFirmian:2026mln} in \eqref{dFV-charges} agree perfectly up to $\mathcal{O}(q^1 y^2)$. 
When comparing the Carter charge, the identity \eqref{FY-pm-KS} proves useful.

\newpage
\bibliographystyle{JHEP}
\bibliography{biblio}

@article{Kim:2023aff,
    author = "Kim, Joon-Hwi and Lee, Sangmin",
    title = "{Symplectic Perturbation Theory in Massive Ambitwistor Space: A Zig-Zag Theory of Massive Spinning Particles}",
    eprint = "2301.06203",
    archivePrefix = "arXiv",
    primaryClass = "hep-th",
    month = "1",
    year = "2023"
}

@article{Kim:2024grz,
    author = "Kim, Joon-Hwi and Kim, Jung-Wook and Lee, Sangmin",
    title = "{Massive twistor worldline in electromagnetic fields}",
    eprint = "2405.17056",
    archivePrefix = "arXiv",
    primaryClass = "hep-th",
    doi = "10.1007/JHEP08(2024)080",
    journal = "JHEP",
    volume = "08",
    pages = "080",
    year = "2024"
}

@article{Kim:2024svw,
    author = "Kim, Joon-Hwi and Kim, Jung-Wook and Kim, Sungsoo and Lee, Sangmin",
    title = "{Classical eikonal from Magnus expansion}",
    eprint = "2410.22988",
    archivePrefix = "arXiv",
    primaryClass = "hep-th",
    doi = "10.1007/JHEP01(2025)111",
    journal = "JHEP",
    volume = "01",
    pages = "111",
    year = "2025"
}

@article{Kim:2025olv,
    author = "Kim, Sungsoo and Lee, Hojin and Lee, Sangmin",
    title = "{Classical eikonal in relativistic scattering}",
    eprint = "2509.01922",
    archivePrefix = "arXiv",
    primaryClass = "hep-th",
    reportNumber = "KIAS-P25044",
    doi = "10.1007/JHEP11(2025)032",
    journal = "JHEP",
    volume = "11",
    pages = "032",
    year = "2025"
}

@article{Gonzo:2024zxo,
    author = "Gonzo, Riccardo and Shi, Canxin",
    title = "{Scattering and Bound Observables for Spinning Particles in Kerr Spacetime with Generic Spin Orientations}",
    eprint = "2405.09687",
    archivePrefix = "arXiv",
    primaryClass = "hep-th",
    doi = "10.1103/PhysRevLett.133.221401",
    journal = "Phys. Rev. Lett.",
    volume = "133",
    number = "22",
    pages = "221401",
    year = "2024"
}

@article{Arkani-Hamed:2019ymq,
    author = "Arkani-Hamed, Nima and Huang, Yu-tin and O'Connell, Donal",
    title = "{Kerr black holes as elementary particles}",
    eprint = "1906.10100",
    archivePrefix = "arXiv",
    primaryClass = "hep-th",
    reportNumber = "NCTS-TH/1905",
    doi = "10.1007/JHEP01(2020)046",
    journal = "JHEP",
    volume = "01",
    pages = "046",
    year = "2020"
}

@article{Carter:1968rr,
    author = "Carter, Brandon",
    title = "{Global structure of the Kerr family of gravitational fields}",
    doi = "10.1103/PhysRev.174.1559",
    journal = "Phys. Rev.",
    volume = "174",
    pages = "1559--1571",
    year = "1968"
}

@article{Alessio:2025flu,
    author = "Alessio, Francesco and Gonzo, Riccardo and Shi, Canxin",
    title = "{Dirac brackets for classical radiative observables}",
    eprint = "2506.03249",
    archivePrefix = "arXiv",
    primaryClass = "hep-th",
    month = "6",
    year = "2025"
}

@article{Akpinar:2025tct,
    author = "Akpinar, Dogan and Brown, Graham R. and Gonzo, Riccardo and Zeng, Mao",
    title = "{Unexpected Symmetries of Kerr Black Hole Scattering}",
    eprint = "2508.10761",
    archivePrefix = "arXiv",
    primaryClass = "hep-th",
    month = "8",
    year = "2025"
}

@article{Compere:2023alp,
    author = "Comp{\`e}re, Geoffrey and Druart, Adrien and Vines, Justin",
    title = "{Generalized Carter constant for quadrupolar test bodies in Kerr spacetime}",
    eprint = "2302.14549",
    archivePrefix = "arXiv",
    primaryClass = "gr-qc",
    doi = "10.21468/SciPostPhys.15.6.226",
    journal = "SciPost Phys.",
    volume = "15",
    number = "6",
    pages = "226",
    year = "2023"
}

@article{Rosquist:2007uw,
    author = "Rosquist, Kjell and Bylund, Tomas and Samuelsson, Lars",
    title = "{Carter's constant revealed}",
    eprint = "0710.4260",
    archivePrefix = "arXiv",
    primaryClass = "gr-qc",
    doi = "10.1142/S0218271809014546",
    journal = "Int. J. Mod. Phys. D",
    volume = "18",
    pages = "429--434",
    year = "2009"
}

@article{Will:2008ys,
    author = "Will, Clifford M.",
    title = "{Carter-like constants of the motion in Newtonian gravity and electrodynamics}",
    eprint = "0812.0110",
    archivePrefix = "arXiv",
    primaryClass = "gr-qc",
    doi = "10.1103/PhysRevLett.102.061101",
    journal = "Phys. Rev. Lett.",
    volume = "102",
    pages = "061101",
    year = "2009"
}

@article{Compere:2021kjz,
    author = "Comp{\`e}re, Geoffrey and Druart, Adrien",
    title = "{Complete set of quasi-conserved quantities for spinning particles around Kerr}",
    eprint = "2105.12454",
    archivePrefix = "arXiv",
    primaryClass = "gr-qc",
    doi = "10.21468/SciPostPhys.12.1.012",
    journal = "SciPost Phys.",
    volume = "12",
    number = "1",
    pages = "012",
    year = "2022"
}

@article{Lynden-Bell:2002dvr,
    author = "Lynden-Bell, Donald",
    title = "{A magic electromagnetic field}",
    eprint = "astro-ph/0207064",
    archivePrefix = "arXiv",
    month = "7",
    year = "2002"
}

@article{Lynden-Bell:2004coe,
    author = "Lynden-Bell, D.",
    title = "{Electromagnetic magic: The Relativistically rotating disk}",
    eprint = "gr-qc/0410109",
    archivePrefix = "arXiv",
    doi = "10.1103/PhysRevD.70.105017",
    journal = "Phys. Rev. D",
    volume = "70",
    pages = "105017",
    year = "2004"
}

@article{Vines:2017hyw,
    author = "Vines, Justin",
    title = "{Scattering of two spinning black holes in post-Minkowskian gravity, to all orders in spin, and effective-one-body mappings}",
    eprint = "1709.06016",
    archivePrefix = "arXiv",
    primaryClass = "gr-qc",
    doi = "10.1088/1361-6382/aaa3a8",
    journal = "Class. Quant. Grav.",
    volume = "35",
    number = "8",
    pages = "084002",
    year = "2018"
}

@article{Chung:2019yfs,
    author = "Chung, Ming-Zhi and Huang, Yu-Tin and Kim, Jung-Wook",
    title = "{Kerr-Newman stress-tensor from minimal coupling}",
    eprint = "1911.12775",
    archivePrefix = "arXiv",
    primaryClass = "hep-th",
    reportNumber = "NCTS-TH/1910",
    doi = "10.1007/JHEP12(2020)103",
    journal = "JHEP",
    volume = "12",
    pages = "103",
    year = "2020"
}

@article{Damgaard:2022jem,
    author = "Damgaard, Poul H. and Hoogeveen, Jitze and Luna, Andres and Vines, Justin",
    title = "{Scattering angles in Kerr metrics}",
    eprint = "2208.11028",
    archivePrefix = "arXiv",
    primaryClass = "hep-th",
    doi = "10.1103/PhysRevD.106.124030",
    journal = "Phys. Rev. D",
    volume = "106",
    number = "12",
    pages = "124030",
    year = "2022"
}

@article{Gibbons:1993ap,
    author = "Gibbons, G. W. and Rietdijk, R. H. and van Holten, J. W.",
    title = "{SUSY in the sky}",
    eprint = "hep-th/9303112",
    archivePrefix = "arXiv",
    reportNumber = "NIKHEF-H93-04, DAMTP-R-92-43",
    doi = "10.1016/0550-3213(93)90472-2",
    journal = "Nucl. Phys. B",
    volume = "404",
    pages = "42--64",
    year = "1993"
}

@article{Ramond:2024ozy,
    author = "Ramond, Paul",
    title = "{On the integrability of extended test body dynamics around black holes}",
    eprint = "2402.02670",
    archivePrefix = "arXiv",
    primaryClass = "gr-qc",
    doi = "10.1088/1361-6382/adb197",
    journal = "Class. Quant. Grav.",
    volume = "42",
    number = "6",
    pages = "065019",
    year = "2025"
}

@article{deFirmian:2026mln,
    author = "de Firmian, Christopher and Vines, Justin",
    title = {{Generalized Carter {\&} R{\"u}diger Constants of $\sqrt{\text{Kerr}}$}},
    eprint = "2602.18790",
    archivePrefix = "arXiv",
    primaryClass = "gr-qc",
    month = "2",
    year = "2026"
}

@article{Kim:2026bqi,
    author = "Kim, Joon-Hwi and Lee, Sangmin",
    title = "{Universality in Relativistic Spinning Particle Models}",
    eprint = "2603.27353",
    archivePrefix = "arXiv",
    primaryClass = "hep-th",
    reportNumber = "CALT-TH 2026-008; KIAS-P26006",
    month = "3",
    year = "2026"
}

@article{Mathisson:1937zz,
    author = "Mathisson, Myron",
    title = "{Neue mechanik materieller systemes}",
    journal = "Acta Phys. Polon.",
    volume = "6",
    pages = "163--200",
    year = "1937"
}

@article{Papapetrou:1951pa,
    author = "Papapetrou, Achille",
    title = "{Spinning test particles in general relativity. 1.}",
    doi = "10.1098/rspa.1951.0200",
    journal = "Proc. Roy. Soc. Lond. A",
    volume = "209",
    pages = "248--258",
    year = "1951"
}

@article{Dixon:1970zza,
    author = "Dixon, W. G.",
    title = "{Dynamics of extended bodies in general relativity. I. Momentum and angular momentum}",
    doi = "10.1098/rspa.1970.0020",
    journal = "Proc. Roy. Soc. Lond. A",
    volume = "314",
    pages = "499--527",
    year = "1970"
}

@article{Kim:2025gis,
    author = "Kim, Jung-Wook and Patil, Raj and Scheopner, Trevor and Steinhoff, Jan",
    title = "{Magnusian: relating the eikonal phase, the on-shell action, and the scattering generator}",
    eprint = "2511.05649",
    archivePrefix = "arXiv",
    primaryClass = "hep-th",
    reportNumber = "CERN-TH-2025-226, HU-EP-25/37-RTG",
    doi = "10.1007/JHEP03(2026)241",
    journal = "JHEP",
    volume = "03",
    pages = "241",
    year = "2026"
}

@article{Kim:2025sey,
    author = "Kim, Joon-Hwi",
    title = "{Manifest symplecticity in classical scattering}",
    eprint = "2511.07387",
    archivePrefix = "arXiv",
    primaryClass = "hep-th",
    reportNumber = "CALT-TH 2025-035",
    month = "11",
    year = "2025"
}

@article{Bern:2023ccb,
    author = "Bern, Zvi and Herrmann, Enrico and Roiban, Radu and Ruf, Michael S. and Smirnov, Alexander V. and Smirnov, Vladimir A. and Zeng, Mao",
    title = "{Conservative Binary Dynamics at Order {\ensuremath{\alpha}}5 in Electrodynamics}",
    eprint = "2305.08981",
    archivePrefix = "arXiv",
    primaryClass = "hep-th",
    doi = "10.1103/PhysRevLett.132.251601",
    journal = "Phys. Rev. Lett.",
    volume = "132",
    number = "25",
    pages = "251601",
    year = "2024"
}

@article{Bargmann:1959gz,
    author = "Bargmann, V. and Michel, Louis and Telegdi, V. L.",
    editor = "Damour, Thibault and Todorov, Ivan and Zhilinskii, Boris",
    title = "{Precession of the polarization of particles moving in a homogeneous electromagnetic field}",
    doi = "10.1103/PhysRevLett.2.435",
    journal = "Phys. Rev. Lett.",
    volume = "2",
    pages = "435--436",
    year = "1959"
}

@article{Newman:1965tw,
    author = "Newman, E. T. and Janis, A. I.",
    title = "{Note on the Kerr spinning particle metric}",
    doi = "10.1063/1.1704350",
    journal = "J. Math. Phys.",
    volume = "6",
    pages = "915--917",
    year = "1965"
}

@article{rudiger:1981rsa,
    author = {Rüdiger, R.},
    title = {Conserved quantities of spinning test particles in general relativity. I},
    journal = {Proceedings of the Royal Society of London. A. Mathematical and Physical Sciences},
    volume = {375},
    number = {1761},
    pages = {185-193},
    year = {1981},
    month = {03},
    issn = {0080-4630},
    doi = {10.1098/rspa.1981.0046}
}

@article{rudiger:1983rsa,
    author = {Rüdiger, R.},
    title = {Conserved quantities of spinning test particles in general relativity. II},
    journal = {Proceedings of the Royal Society of London. A. Mathematical and Physical Sciences},
    volume = {385},
    number = {1788},
    pages = {229-239},
    year = {1983},
    month = {01},
    issn = {0080-4630},
    doi = {10.1098/rspa.1983.0012}
}

@article{Kim:2025xka,
    author = "Kim, Joon-Hwi",
    title = "{Note on the Kerr Spinning-Particle Equations of Motion}",
    eprint = "2512.23697",
    archivePrefix = "arXiv",
    primaryClass = "gr-qc",
    reportNumber = "CALT-TH 2025-006",
    month = "12",
    year = "2025"
}

@article{Kim:2026opo,
    author = "Kim, Joon-Hwi",
    title = "{The Kerr two-twistor particle}",
    eprint = "2602.19495",
    archivePrefix = "arXiv",
    primaryClass = "gr-qc",
    reportNumber = "CALT-TH 2026-001",
    month = "2",
    year = "2026"
}

@article{Kim:2026yqo,
    author = "Kim, Joon-Hwi",
    title = "{The Kerr-Newman two-twistor particle}",
    eprint = "2603.07537",
    archivePrefix = "arXiv",
    primaryClass = "gr-qc",
    reportNumber = "CALT-TH 2026-011",
    month = "3",
    year = "2026"
}

@article{Witzany:2019nml,
    author = "Witzany, Vojt{\v{e}}ch",
    title = "{Hamilton-Jacobi equation for spinning particles near black holes}",
    eprint = "1903.03651",
    archivePrefix = "arXiv",
    primaryClass = "gr-qc",
    doi = "10.1103/PhysRevD.100.104030",
    journal = "Phys. Rev. D",
    volume = "100",
    number = "10",
    pages = "104030",
    year = "2019"
}

@article{Drummond:2023loz,
    author = "Drummond, Lisa V. and Hanselman, Alexandra G. and Becker, Devin R. and Hughes, Scott A.",
    title = "{Extreme mass-ratio inspiral of a spinning body into a Kerr black hole I: Evolution along generic trajectories}",
    eprint = "2305.08919",
    archivePrefix = "arXiv",
    primaryClass = "gr-qc",
    month = "5",
    year = "2023"
}

@article{Skoupy:2024uan,
    author = "Skoup{\'y}, Viktor and Witzany, Vojt{\v{e}}ch",
    title = "{Analytic Solution for the Motion of Spinning Particles in Kerr Spacetime}",
    eprint = "2411.16855",
    archivePrefix = "arXiv",
    primaryClass = "gr-qc",
    doi = "10.1103/PhysRevLett.134.171401",
    journal = "Phys. Rev. Lett.",
    volume = "134",
    number = "17",
    pages = "171401",
    year = "2025"
}

@article{Skoupy:2025nie,
    author = "Skoup{\'y}, Viktor and Piovano, Gabriel Andres and Witzany, Vojt{\v{e}}ch",
    title = "{Spherical inspirals of spinning bodies into Kerr black holes}",
    eprint = "2506.20726",
    archivePrefix = "arXiv",
    primaryClass = "gr-qc",
    doi = "10.1103/x9yy-c2jq",
    journal = "Phys. Rev. D",
    volume = "112",
    number = "12",
    pages = "124054",
    year = "2025"
}

@article{Piovano:2025aro,
    author = "Piovano, Gabriel Andres",
    title = "{Particles with precessing spin in Kerr spacetime: Analytic solutions for eccentric orbits and homoclinic motion near the equatorial plane}",
    eprint = "2510.09597",
    archivePrefix = "arXiv",
    primaryClass = "gr-qc",
    doi = "10.1103/jzbw-m1cp",
    journal = "Phys. Rev. D",
    volume = "113",
    number = "6",
    pages = "064024",
    year = "2026"
}

@article{Newman:1974fr,
    author = "Newman, E. T. and Winicour, J.",
    title = "{A curiosity concerning angular momentum}",
    doi = "10.1063/1.1666761",
    journal = "J. Math. Phys.",
    volume = "15",
    pages = "1113--1115",
    year = "1974"
}

@article{Kim:2024mpy,
    author = "Kim, Joon-Hwi",
    title = "{Newman-Janis Algorithm from Taub-NUT Instantons}",
    eprint = "2412.19611",
    archivePrefix = "arXiv",
    primaryClass = "gr-qc",
    reportNumber = "CALT-TH 2024-051",
    month = "12",
    year = "2024"
}

@article{Adamo:2023fbj,
    author = "Adamo, Tim and Bogna, Giuseppe and Mason, Lionel and Sharma, Atul",
    title = "{Scattering on self-dual Taub-NUT}",
    eprint = "2309.03834",
    archivePrefix = "arXiv",
    primaryClass = "hep-th",
    doi = "10.1088/1361-6382/ad12ee",
    journal = "Class. Quant. Grav.",
    volume = "41",
    number = "1",
    pages = "015030",
    year = "2024"
}

@article{Kim:2024dxo,
    author = "Kim, Joon-Hwi",
    title = "{Single Kerr-Schild metric for Taub-NUT instanton}",
    eprint = "2405.09518",
    archivePrefix = "arXiv",
    primaryClass = "hep-th",
    reportNumber = "CALT-TH 2024-020",
    doi = "10.1103/PhysRevD.111.L021703",
    journal = "Phys. Rev. D",
    volume = "111",
    number = "2",
    pages = "L021703",
    year = "2025"
}

@article{Guevara:2020xjx,
    author = "Guevara, Alfredo and Maybee, Ben and Ochirov, Alexander and O'connell, Donal and Vines, Justin",
    title = "{A worldsheet for Kerr}",
    eprint = "2012.11570",
    archivePrefix = "arXiv",
    primaryClass = "hep-th",
    doi = "10.1007/JHEP03(2021)201",
    journal = "JHEP",
    volume = "03",
    pages = "201",
    year = "2021"
}

@article{Arkani-Hamed:2017jhn,
    author = "Arkani-Hamed, Nima and Huang, Tzu-Chen and Huang, Yu-tin",
    title = "{Scattering amplitudes for all masses and spins}",
    eprint = "1709.04891",
    archivePrefix = "arXiv",
    primaryClass = "hep-th",
    reportNumber = "NCTS-TH/1714, NCTS-TH-1714",
    doi = "10.1007/JHEP11(2021)070",
    journal = "JHEP",
    volume = "11",
    pages = "070",
    year = "2021"
}

@article{Guevara:2018wpp,
    author = "Guevara, Alfredo and Ochirov, Alexander and Vines, Justin",
    title = "{Scattering of Spinning Black Holes from Exponentiated Soft Factors}",
    eprint = "1812.06895",
    archivePrefix = "arXiv",
    primaryClass = "hep-th",
    doi = "10.1007/JHEP09(2019)056",
    journal = "JHEP",
    volume = "09",
    pages = "056",
    year = "2019"
}

@article{Chung:2018kqs,
    author = "Chung, Ming-Zhi and Huang, Yu-Tin and Kim, Jung-Wook and Lee, Sangmin",
    title = "{The simplest massive S-matrix: from minimal coupling to Black Holes}",
    eprint = "1812.08752",
    archivePrefix = "arXiv",
    primaryClass = "hep-th",
    reportNumber = "NCTS-TH/1817",
    doi = "10.1007/JHEP04(2019)156",
    journal = "JHEP",
    volume = "04",
    pages = "156",
    year = "2019"
}

@article{Guevara:2019fsj,
    author = "Guevara, Alfredo and Ochirov, Alexander and Vines, Justin",
    title = "{Black-hole scattering with general spin directions from minimal-coupling amplitudes}",
    eprint = "1906.10071",
    archivePrefix = "arXiv",
    primaryClass = "hep-th",
    doi = "10.1103/PhysRevD.100.104024",
    journal = "Phys. Rev. D",
    volume = "100",
    number = "10",
    pages = "104024",
    year = "2019"
}

@article{Chung:2019duq,
    author = "Chung, Ming-Zhi and Huang, Yu-Tin and Kim, Jung-Wook",
    title = "{Classical potential for general spinning bodies}",
    eprint = "1908.08463",
    archivePrefix = "arXiv",
    primaryClass = "hep-th",
    reportNumber = "NCTS-TH/1907",
    doi = "10.1007/JHEP09(2020)074",
    journal = "JHEP",
    volume = "09",
    pages = "074",
    year = "2020"
}

\end{document}